\documentclass[journal,comsoc]{IEEEtran}
\usepackage[T1]{fontenc}% optional T1 font encoding
\usepackage{cite}
\usepackage{amsmath,amssymb,amsfonts}
\usepackage{algorithmic}
\usepackage{graphicx}
\usepackage{textcomp}
\usepackage{multirow}
\usepackage{hyperref}

\ifCLASSINFOpdf
\else
\fi
\usepackage{amsmath}
\usepackage[cmintegrals]{newtxmath}
\begin{document}
\title{Smart Urban Mobility: When Mobility Systems\\ Meet Smart Data}

\author{Zineb Mahrez,~\IEEEmembership{Student,~IEEE,}
        Essaid Sabir,~\IEEEmembership{Senior,~IEEE,}
        Elarbi Badidi,~\IEEEmembership{Member,~IEEE,}
        Walid Saad,~\IEEEmembership{Fellow,~IEEE,}
        and~Mohamed~Sadik,~\IEEEmembership{Member,~IEEE}% <-this % stops a space
\thanks{Z. Mahrez, E. Sabir, and M. Sadik are with NEST Research Group, ENSEM, Hassan II University of Casablanca, Morocco.}% <-this % stops a space
\thanks{E. Badidi is with the College of Information Technology, UAE University, PO Box. 15551, AL-AIN, UAE.}% <-this % stops a space
\thanks{W. Saad is with Wireless@VT, Electrical and Computer Engineering Department, Virginia Tech, Blacksburg, VA 24061, USA.}% <-this % stops a space
\thanks{Manuscript received May 1st, 2020; revised August 26, 2020.}}


\markboth{IEEE Communications Surveys & Tutorials}%
{Shell \MakeLowercase{\textit{et al.}}: Bare Demo of IEEEtran.cls for IEEE Communications Society Journals}



\maketitle

\begin{abstract}
Cities around the world are expanding dramatically, with urban population growth reaching nearly 2.5 billion people in urban areas and road traffic growth exceeding 1.2 billion cars by 2050. The economic contribution of the transport sector represents 5\% of the GDP in Europe and costs an average of US \$482.05 billion in the United States. These figures indicate the rapid rise of industrial cities and the urgent need to move from traditional cities to smart cities. 
This article provides a survey of different approaches and technologies such as intelligent transportation systems (ITS) that leverage communication technologies to help maintain road users safe while driving, as well as support autonomous mobility through the optimization of control systems. The role of ITS is strengthened when combined with accurate artificial intelligence models that are built to optimize urban planning, analyze crowd behavior and predict traffic conditions. AI-driven ITS is becoming possible thanks to the existence of a large volume of mobility data generated by billions of users through their use of new technologies and online social media. The optimization of urban planning enhances vehicle routing capabilities and solves traffic congestion problems, as discussed in this paper. From an ecological perspective, we discuss the measures and incentives provided to foster the use of mobility systems. We also underline the role of the political will in promoting open data in the transport sector, considered as an essential ingredient for developing technological solutions necessary for cities to become healthier and more sustainable.

\end{abstract}


\footnote{This work has been conducted within the framework of the \textit{Mobicity} project funded by the Moroccan Ministry of Higher Education and Scientific Research, and the National Centre for Scientific and Technical Research.\\
W. Saad was supported by the U.S. National Science Foundation under Grant IIS-1633363.}

\markboth
{Smart Urban Mobility: When Mobility Systems Meet Smart Data}{Z. Mahrez \headeretal}
%%{Smart Urban Mobility: When Smart Data Meets Mobility Systems}{Z. Mahrez \headeretal}

% \fi
\IEEEpeerreviewmaketitle

\section{INTRODUCTION}

Effective transportation systems contribute to economic growth and the building of prosperous cities. For example, Chicago, Singapore, Beijing, London and many other cities around the world, have become hubs of trade and commerce thanks to their well-established transportation systems \cite{cytron2010role}. The economic contribution of the transport sector represents 5\% of the GDP in Europe \cite{ECGDP2018} and costs an average of US \$482.05 billion in the United States \cite{GDPUSA}.
However, as the urban population is expected to reach 68\% of the world's population by 2050 \cite{UNpopulationprojection}, new cities and new infrastructures, including buildings and transportation systems, need to be built to meet the needs of the growing population. In the United States alone, urban expansion costs about \$400 billion per year \cite{economy2014better}. As a result of this urban expansion, urban mobility will be the main source of pollution to affect the environment severely. Gas emissions from urban transport systems are estimated at around 60\% of the total greenhouse gas emissions \cite{toledo2018urban}. According to INRIX research, US cities are ranked among the 10 most congested cities worldwide, with serious consequences in terms of cost and waiting time for drivers \cite{cookson2018inrix}. For example, in Los Angeles, drivers spend about 102 hours a year in traffic jams, costing the city about \$19.2 billion. In Moscow, drivers spend 91 hours in traffic jams \cite{cookson2018inrix}. In Europe, congestion in urban areas costs around 100 billion euros per year, which also negatively impacts the economy \cite{ECmobility}. In addition, road accidents kill more than 1.2 million people a year \cite{worldhealthorganisation2015}.\\
\noindent In a global context, and in accordance with the sustainable development goals (SDGs) set by the United Nations, all these urban mobility challenges could be addressed through implemented actions that will help build more sustainable cities by 2030.   The SDGs involve all countries and sectors, including the public and private sectors, academia and civil society, to work in an interdisciplinary field related to sustainability to ensure peace and prosperity for all.
In this case, urban mobility is linked to SDG 3 (Health and well-being), SDG7 (affordable and clean energy), SDG 9 (Industry, innovation and infrastructure), SDG 11 (Sustainable cities and communities), SDG 13 (Climate action), and SDG 17 (Partnerships for the goals) \cite{SDGs}.\\
\noindent In the same context, the European Commission (EC) has created an observatory dedicated to urban mobility (http://www.eltis.org/) to facilitate the exchange of information on sustainable urban development. The EC also invited all EU Member States to review urban transport planning at the national level and worked on a White Paper on Transport Policy entitled " Roadmap for a European Area" unique transport" \cite{ec2011white}. 
In addition to urban transport policies, the EC has developed the URBACT (http://www.urbact.eu/) and the CIVITAS (http://www.civitas.eu/) platforms to help European cities build their capacity in sustainable urban mobility. These platforms encourage EU cities to take new initiatives and develop sustainable solutions for urban mobility in order to restructure cities and contribute to their prosperity. Recently, the European Commission has developed new strategies for local governments and has invited them to join the movement of building the smart mobility concept. \emph{Smart mobility}, one of the smart city pillars, uses new technologies to connect users, infrastructures and vehicles in order to improve the speed of transfer in urban areas, reduce congestion, enhance people's safety and reduce air and noise pollution \cite{benevolo2016smart}. In this perspective, each country has generated a specific plan adapted to its urban context. The evolution of smart mobility strategies along with the emergence of IoT have enabled the development of a wide range of technologies that strengthen the current public transportation services, offer adapted services to the needs of users, such as parking, emergency cases, charging, and foster the use of electric vehicles through a variety of incentives.\\
\noindent Several surveys on smart mobility have been published, for example, \cite{faria2017smart} and \cite{benevolo2016smart} provided a theoretical analysis, with a discussion of smart mobility from an ICT perspective, and outlined the different existing technologies that support smart mobility services. The surveys in \cite{al2019smart} and \cite{bayram2014survey} focused respectively on smart parking and electric vehicles, as a specific aspect of smart mobility. Nevertheless, the discussion about smart mobility systems could not avoid artificial intelligence and vehicle routing. Moreover, the implementation of smart mobility vision requires a cross-sector collaboration, i.e., public-private-academic sectors and citizens, to build a foundation of tomorrow's smart mobility. The main contribution of this paper is to provide a holistic analysis of the existing literature on smart mobility practices. In particular, we present the various mobility systems, including intelligent transportation systems, artificial intelligence-driven smart mobility systems, vehicle routing systems, and e-mobility systems, that form a package of optimisation tools that can be used to implement the smart mobility vision. We also showcase the substantial role of open data and analyze its economic value in building intelligent, sustainable, and resilient transportation systems.
The rest of the paper is organized as follows. Section II surveys the different existing ITS systems to ensure safety, real-time services, and assist in autonomous mobility. Section III discusses AI-driven smart mobility systems. Sections IV and V present, respectively, different applications of vehicle routing systems and e-mobility systems. Section VI presents smart mobility from a data perspective, by listing examples of data sources and discussing related data attributes. Section VII raises the problem of technology transfer, particularly in emerging countries. Section VIII discusses the remaining challenges faced by smart mobility technologies. Finally, section IX concludes the paper and highlights open research challenges.

\section{Smart mobility and ITS }
This section explores the current literature on the various applications of intelligent transportation systems, with a particular focus on safety services, real-time applications and autonomous mobility.
\noindent Intelligent transportation systems (ITS), represent a form of new technologies that assist in solving urban problems such as traffic congestion, energy consumption, and road safety \cite{isosmartcity2014}.
ITSs enable technological cooperation between vehicles, infrastructures, and users through the exchange of information on traffic conditions. The interactions created between these elements are classified into several communication models, i.e., vehicle-to-infrastructure (V2I), vehicle-to-vehicle (V2V), vehicle-to-pedestrian (V2P), and vehicle-to-everything (V2X). V2I allows vehicles to communicate with the road infrastructure via cellular networks, onboard units (OBUs), and roadside units (RSUs). It is used in toll stations, parking payment, and alert systems to warn drivers of safety risks. Municipal officials use V2I to improve road design, reduce congestion and manage the road network efficiently by providing critical information to road users. For instance, to address the inefficiency of current traffic control systems, the authors of  \cite{bravo2016smart} and \cite{al2015applications} implemented decision support systems, using big data applications to complement and optimize the role of traffic lights, thereby reducing travel time and pollution. This approach reduced traffic congestion and wait times for drivers, reducing gas emissions by 20\%. V2V communication enables inter-vehicle communication via wireless channels and OBUs to improve road safety and manage congestion. V2V is used in several situations; for instance, it alerts drivers when the vehicle ahead slows down; it is also used to manage traffic lights according to the current traffic flow and facilitates cooperative driving to avoid accidents \cite{guo2015integration}. Moreover, some ITS applications may be a good substitute for traffic lights when controlling intersections and traffic planning. They collect information on mobility (speed, vehicle position) through V2I or V2V communications over wireless links, and use groups of vehicles to reduce waiting times and improve fairness compared to traditional traffic lights \cite{cheng2017fuzzy}. Also, V2X represent a generalization of V2I and V2V, and V2P consists of transmitting safety signals from pedestrians mobile applications to vehicles to avoid collisions and accidents. Furthermore, the information exchange requires the application of adequate communication protocols, such as the IEEE 802.11p standard, IEEE 802.15.4/ZigBee, LTE-V2V, Bluetoooth, and cellular technologies \cite{arena2019overview}. However, some communication protocols are facing technical challenges during data processing, such as communication network congestion, and computational overload, which are due to the massive amount of heterogeneous data collected from smart sensors. These challenges may result in high delays in system decisions, and may cause traffic congestion or road accidents, especially when it comes to supporting autonomous vehicles. To address these technical problems, several works \cite{zeng2019jointplatoon},\cite{saad2019vision},\cite{ferdowsi2019deep}, focused on the cellular communication network (5G) to support V2X communication in providing low latency (from $1$~ms to $2$~ms), security and privacy, high reliability, and optimal cost efficiency. For instance, the authors of \cite{zeng2019jointplatoon} focused on the use of cellular vehicle-to-everything communication (C-V2X) to study the problem of vehicle's control system in order to ensure a successful platooning of wireless connected autonomous vehicles. Furthermore, the authors of \cite{saad2019vision} explore the use of the sixth-generation wireless system (6G), in enabling a 3D visualization of the radio environment, based on its multi-purpose sensing system that delivers multiple services such as, tracking, control, localization, and computing.\\ 
\noindent Next, we discuss concrete examples of ITS applications that use V2I, V2V and V2X communication for safety purposes, real-time services, and autonomous mobility.
 
\subsection{Safety services of ITS}
Road accidents are a major public concern. According to the World Health Organization, more than 1.2 million deaths are recorded each year on the world's roads \cite{worldhealthorganisation2015}. Low and middle-income countries are the most affected, accounting for 90\% of all deaths worldwide. In these countries, vulnerable road users, including pedestrians, cyclists, and motorcyclists, represent 50\% of road victims. For instance, in the United States, 5.3\% of fatal road crashes are related to collisions at road intersections controlled by traffic signs \cite{noble2016influence}. Because of this alarming situation, road safety has been integrated into the new 2030 agenda of the United Nations, which has led countries to implement new policies and laws related to the speed limit, the obligation to use motorcycle helmets and vehicle seatbelts, and drunk driving penalties \cite{worldhealthorganisation2015}. 
In this regard, many urban mobility actors, including manufacturers, public authorities, and academic institutions have committed to exploring the contribution of ITS in improving road safety. At the same time, researchers are increasingly looking for new ITS tools to better manage traffic, especially at intersections. For instance, the work in \cite{hubner2012its} used ITS security notifications to alert drivers about hidden objects, warn vulnerable users of an approaching vehicle, and send automatic collision notification to drivers. To evaluate the effective use of ITS in critical situations, authors in \cite{noble2016influence} conducted an investigation that measures and evaluates the level of user compliance with ITS notifications compared to traditional road instructions. The study revealed that users are more sensitive to the in-vehicle adaptive stopping display than to a traditional stop sign. ITS applications are also used to provide solutions to protect vulnerable road users, i.e., pedestrians, older drivers, and young novice drivers, based on V2P communication and pedestrian-to-infrastructure (P2I) interaction \cite{sumalee2018smarter}. More detailed ITS solutions designed for vulnerable road users are provided in Table \ref{Table 2}. In another context, in case of disasters, ITS services are crucial in providing quick and effective responses. For instance, the work in \cite{hamza2008dynamic} proposed a smart traffic evacuation management system and two algorithms describing the measures to be taken in times of crisis. In the same context, the authors of \cite{song2017deepmob} developed a system for predicting human evacuation following a natural disaster using deep learning techniques. To expand the overview of ITS applications for safety services, table \ref{table:1} presents a non-exhaustive list of ITS technologies applied to prevent intersection collision, and protect drivers and users from adverse weather conditions.\\

\begin{table}[!htb]
\caption{ITS for vulnerable road users \cite{regan2001intelligent}}
\label{Table 2}
\begin{tabular}{ll}
\textbf{Vulnerable road user category} & \textbf{ITS Solution}  \\
\hline
\multirow{3}{*}{\textbf{Young novice drivers}}  & Forward collision warning systems \\  & \begin{tabular}[c]{@{}l@{}}Seat-belt reminder systems\end{tabular}  \\
  & Breath alcohol detection\\
  \hline
\multirow{7}{*}{\textbf{Older drivers}} & Mayday systems\\& Vision enhancement systems\\  & \begin{tabular}[c]{@{}l@{}}Rear collision warning systems and \\ in-vehicle navigation systems\end{tabular} \\& Lane departure warning \\& \begin{tabular}[c]{@{}l@{}}Lane change collision \\ warning systems \end{tabular}\\
\hline
\multirow{3}{*}{\textbf{Pedestrians and bicyclists}} & Speed alerting\\& Variable speed warning\\ & Vision enhancement systems\\
\hline
\end{tabular}
\end{table}

\begin{table*}
\ref{table:1}
\caption{{Non exhaustive list of ITS Services for safety } \cite{dennyv2isafety,stevens2012benefits,richard2015multiple,valldorf2008advanced,itssafetysolutions}.}
\label{table:1}
 \begin{tabular}{|l|c|c|c|}
 \hline
\textbf{ITS Technology}&\textbf{Description} & \textbf{Type of technology} & \textbf{Sensors}\\
\hline
\multicolumn{4}{|c|}{\textbf{Prevent intersection collision}}   \\
\hline
Curve speed warning &	 \begin{tabular}[c]{@{}c@{}} - Warning when travel speed \\ exceeds a curve safe speed \\ - Prevents rollover, run-off-road \end{tabular} & Roadway safety systems & Radar\\
\hline
Signalized left turn assist	&\begin{tabular}[c]{@{}c@{}}  - Assisting left turns when visibility \\ is limited at intersections\\ - Prevents vehicle head-on and sideswipe \end{tabular}& Vehicle safety systems & Yaw-rate sensor\\
\hline
Red light violation warning &	\begin{tabular}[c]{@{}c@{}} - Indicates a potential for violating a red light \\ - Multivehicle collision including sideswipe or\\  broadside accident and rear-end crashes \end{tabular} & Vehicle safety systems &Radar, sonar\\
\hline
\multicolumn{4}{|c|}{\textbf{Adverse weather conditions}}   \\
\hline
Spot weather information warning & \begin{tabular}[c]{@{}c@{}} Alerting drivers about adverse weather \\ high wind, flooding \end{tabular} & Roadway safety systems & RWIS\\
\hline
Intersection collision avoidance systems &  \begin{tabular}[c]{@{}c@{}} Warning about \\ approaching dangerous cross traffic  \end{tabular}   & Roadway safety systems & Radar\\
\hline 
Hazardous location notification & \begin{tabular}[c]{@{}c@{}}  Warning about possible\\ hazardous locations ahead \end{tabular} & Roadway safety systems & Rollover sensor\\
\hline 
Wildlife detection systems & Identifying animals approaching the roadway & Roadway safety systems & Infrared\\
\hline 
Low Bridge Warning & Alerting approaching vehicles & Roadway safety systems & GPS\\
\hline
Advanced braking assistance & Performs a full brake when an obstacle is detected  & Vehicle safety systems & Laser, Radar\\
\hline
Electronic stability control & Detects and reduces the risk to skid & Vehicle safety systems & Wheel-speed, yaw-rate\\
\hline
Intelligent Speed Adaptation & \begin{tabular}[c]{@{}c@{}} Warning when the vehicle \\ exceeds the speed limit  \end{tabular}  & Vehicle safety systems & GPS \\
\hline
Wrong way driving warning & The vehicle is out of short range & Vehicle safety systems & GPS\\
\hline
Lane departure warning systems &  Alerting vehicle drifting  &  Vehicle safety systems & Cameras\\
\hline
Drowsy driver warning systems &  Alerting about signs of fatigue in eye's driver & Vehicle safety systems & Video\\
\hline Vulnerable road user warning & \begin{tabular}[c]{@{}c@{}} Warning of the presence of vulnerable \\ road users on the roadway  \end{tabular} & Vehicle safety systems & Radar\\
\hline
Automatic crash notification systems & \begin{tabular}[c]{@{}c@{}} Transmitting information on the \\ vehicle's location to a call center \end{tabular} & Emergency response & On-board sensor\\
\hline 
Emergency vehicle preemption & Providing right-of-way to emergency vehicles & Emergency response & GPS, radio transmitter\\
\hline

\end{tabular} 
  
\end{table*}

\noindent Moreover, real-time service represents a key parameter for maintaining safety in urban mobility. The second part of this section discusses ITS applications that use real-time technologies.

\subsection{Real-time ITS services}

Communicating the right information to drivers in the right place at the right time is extremely important. To this end, several ITS applications have been deployed to provide real-time services to drivers. For instance, the work in \cite{hoong2012road} proposed a bayesian network model that relies on mobile devices to collect infrastructure data, such as maps and road design, and data from tweets, on accidents, road construction or roadblock. The model analyzes contextual data, including direction, speed, altitude, driver information, and vehicle type, to assess traffic conditions in real-time and predict traffic jams. Other works relied on traffic visualization that is based on reconstructing traffic conditions according to current traffic data, to visually display current road and traffic conditions. For instance, the work in \cite{li2017city} focused on mobile sensor data (GPS data) to estimate the real-world traffic conditions. The estimation process enabled reconstruction, visualization, and animation of traffic in 2D and 3D using a statistical learning method. The authors also addressed the problem of data coverage in some city areas by using a metamodel based simulation to complete the missing data. In \cite{lin2012service}, the authors designed a framework to address real-time dynamic traffic congestion issues by injecting instantaneous data and supporting prediction strategies for traffic management. Their work consists of integrating data-driven dynamic application systems (DDDAS) with service-oriented architecture (SOA) and Web services technology, for a real-time decentralized traffic signal control system. The authors used real-time traffic information from each intersection to determine the necessary service data and create a data-driven dynamic tree. Traffic signal control systems typically calculate three factors, cycle time, division rate and offset, to regulate traffic flow. The authors devised a control strategy in the prediction module to manipulate these three factors, and they used web services technology to support data exchange and communication.\\ 
However, at some locations in the city, the actual positioning infrastructure may lead to an erroneous position that can affect the quality of the information provided by ITS services. The works in \cite{clausen2017assessment} \cite{drawil2013gps} were conducted to address this problem, by evaluating the positioning performance of the global navigation satellite system (GNSS) and classifying the geospatial positioning accuracy. In addition, many products and prototypes, such as the virtual dynamic message signals (VDMS) system, were designed to provide real-time traffic information to users \cite{ma2016virtual}. The authors in \cite{chmiel2016insigma} also implemented an intelligent transportation system, called INSIGMA that provides information on traffic intensity as well as road transport infrastructure and warns of possible dangerous events. Moreover, ITS applications can use GPS satellites to track the progress of a vehicle to a target point in real-time. Such a tracking system also provides other types of instant vehicle information such as speed, route, and stopping points. The authors of \cite{darwish2017empowering}  developed a system for detecting and tracking vehicles from camera footages to determine the number of moving vehicles in a series of images. Table \ref{Table 4} lists the different models used by real-time services for tracking. 

\begin{table}
\caption{Models for tracking arrival time prediction \cite{altinkaya2013urban}}
\label{Table 4}
\begin{tabular}{ll}
\textbf{Models} & \textbf{Description}\\
\hline
Time series models & \begin{tabular}[c]{@{}l@{}}- Use historical data about traffic patterns\\ - There is a risk of prediction \\ inaccuracy when the relationship between\\ historical data and real-time \\data is changing \end{tabular} \\
\hline
Regression models& \begin{tabular}[c]{@{}l@{}}- Use a set of independent inputs \\e.g: distance, number of stops,\\ boarding and alighting\\ passengers and weather descriptors\\ - The application of this model \\ is limited in highly inter-correlated \\transportation systems  \end{tabular} \\
\hline
Kalman filtering model & \begin{tabular}[c]{@{}l@{}} - Relies on real-time location data and \\statistical dynamic travel time estimation.\\- The data could be obtained from \\automatic vehicle location (AVL) \\and automatic passenger counter (APC)\\mounted on urban buses for instance \end{tabular}\\
\hline
Artificial neural network & \begin{tabular}[c]{@{}l@{}}  - It   focuses on the
complex \\non-linear relationship between \\travel time and the independent variables.\\- It is gaining popularity in predicting time\\ arrival and outperforms other algorithms.\end{tabular}\\
\hline
Hybrid models & \begin{tabular}[c]{@{}l@{}} - Integrate two or more above-mentioned \\models to improve estimation \\precision and prediction accuracy\end{tabular}\\
\hline
\end{tabular}
\end{table}

\noindent The concept of crowdsourcing also contributes to improving the accuracy of the vehicle positioning system, where real-time tracking systems have been deployed to schedule vehicle departure and arrival times and more specifically for bus tracking and arrival time prediction. \cite{mukheja2017smartphone}. ITS real-time services are also widely used in the emergency medicine supply chain to meet legislative requirements for tracking information on transported and dispensed drugs while meeting traceability, secure delivery requirements and proper handling \cite{zuazola2013telematics}. Moreover, ITS real-time services are excellent solutions for parking areas. For example, in Vienna, a real-time model based on existing data, such as electronic parking tickets, counts of car parks, traffic flow data, was developed to provide an accurate estimate of the occupation of the parking area \cite{hossinger2014development}. Other works focused on the guidance to the closest parking areas, for instance, the authors in \cite{coulibaly2018development} developed a mobile application that uses a demonstrator based on RFID. The application service allows users to find free parking spaces in real-time, book parking spaces in advance and pay online. In the same context, SmartyPark is also a parking management solution that provides real-time location and guidance to the nearest available parking location using a Senstenna-based mobile application, a 5th generation sensorless technology that can detect objects. The solution is a desktop application designed to provide the parking manager with a real-time view of the parking occupancy. The application allows the user to make payment and booking online and provide a "watch over my car" service in real-time \cite{smartilabsmartypark}.\\  
\noindent The next part of this section discusses the vital role of the two previously cited ITS services (response-time and safety applications), in enhancing latency, reliability and wireless networking performance, when dealing with autonomous vehicles.

\subsection{ITS supporting Autonomous mobility}
The idea of autonomous vehicles (AV) was born to face the problem of human errors likely to cause accidents. Nearly 90\% of road accidents are caused by human errors \cite{Policyvictoria}.  Meanwhile, the transportation ecosystem is witnessing the emergence of automated driving solutions to assist users, reduce congestion and vehicle emissions, eliminate traffic delays and, in particular, improve people's accessibility \cite{Bagloee2016}.
A standard classification of automation levels based on ITS has been developed to offer fully or partially automatic functionality to help drivers make decisions and act instantly in critical situations, as detailed below:
\begin{itemize}
\item \textbf{Manual driving}: the vehicle is entirely controlled by the driver, who performs all driving tasks (steering, braking, throttling, and shifting).
\item \textbf{Assisted driving}: Integration of a single automated function, such as adaptive cruise control (ACC), which adjusts the vehicle speed to maintain a safe distance from the leading vehicles.
\item \textbf{Semi-automated driving}: Integration of intelligent vehicle technologies for control functions, for instance, adaptive cruise control associated with track centering.
\item \textbf{Highly automated driving}: The control functions are highly automated. However, the driver can intervene to take over when necessary.
\item \textbf{Fully automated driving}: the vehicle is intelligent and can comprehensively perform all driving tasks (reading traffic signs, object detection at the front by radar sensors, traffic light detection).
\end{itemize}
More examples of monitoring technologies for AVs services are provided in Table \ref{Table 3}.\\
AVs are expected to represent 30\% of vehicles in 2040, which could create fierce competition between car manufacturers. Today, many automakers (BMW, Mercedes Benz, Nissan, General Motors) and Google are developing prototypes of autonomous vehicles. However, some AV cars have already recorded minor road accidents, for example, a Google car collided with a public bus in Silicon Valley \cite{Bagloee2016}. The risks could be attributed to several factors, including issues related to the connectivity, autonomy, sensing, limited data processing capabilities and also problems related to current data analysis infrastructures, which can lead to some data processing problems such as high latency problems. To improve the autonomous attitude of AVs, and overcome ITS reliability and latency, the authors in \cite{ferdowsi2019deep} investigated the potential of exploiting deep learning techniques together with connected vehicle technologies. The study consisted of implementing an edge centric solution combined with deep learning techniques that seek to improve data analytics capabilities, thus enhancing ITS latency and reliability. The system involves collecting raw heterogeneous sets of data related to the environment of AVs that are analyzed and processed at the vehicle or roadside smart sensor level. The processed data provide appropriate measures accordingly, to better monitor road conditions, such as lane change, acceleration or speed reduction.  Moreover, the performance of ITS is highly required in the context of wireless connected autonomous vehicular platoons. Vehicle platooning refers to a group of vehicles, that are operating collaboratively, and are arranged to form a leader-follower model. The work in \cite{zeng2019jointplatoon} addressed the problem of stability and reliability of wireless connected vehicular platoons, by developing a framework that jointly analyzes both communication and control systems. In practical terms, all vehicles within the platoon are interacting through V2V communication. For instance, the lead vehicle exchanges information with the following vehicles about target speed, velocity and acceleration. The following vehicles coordinate accordingly using cooperative adaptive cruise control (CACC), to ensure the stability of platooning and control inter-vehicle distances. In the same perspective, authors in \cite{zeng2019joint} adopted a similar approach, i.e., a joint communication and control framework, to study the path tracking problem for connected and autonomous vehicles (CAV). To prevent the instability due to a delayed transmission of information, the framework enables CAV to exploit V2X communication, to sense and track environment such as location, straight lines, heading angle and circular curves, to be able to determine its navigation path and adjust the control system accordingly.\\
\noindent In view of technological advances discussed and adopted in the autonomous mobility landscape, the arrival of autonomous vehicles on the market is expected to reduce several costs related to accidents, environmental degradation and transit tickets. In this perspective, several universities are trying to strengthen links between city officials and researchers to better prepare for the advent of autonomous vehicles \cite{Bagloee2016} \cite{brainonboard}.

\begin{table}
\caption{Automation levels attributes \cite{Bagloee2016} \cite{brainonboard}}
\label{Table 3}
\begin{tabular}{ll}
\textbf{AV Services} & \textbf{\begin{tabular}[c]{@{}l@{}}Monitoring   technologies\end{tabular}} \\
\hline
\begin{tabular}[c]{@{}l@{}}Forward collision   warnings (FCW)\end{tabular} & Camera, radar \\
\hline
\begin{tabular}[c]{@{}l@{}}Lane departure  warnings\end{tabular} &
\begin{tabular}[c]{@{}l@{}}Video, laser,\\ Infrared sensor\end{tabular} \\
\hline
\begin{tabular}[c]{@{}l@{}}Electronic stability   control\end{tabular} & \begin{tabular}[c]{@{}l@{}}Wheel speed sensor,\\ steering angle\end{tabular} \\
\hline
\begin{tabular}[c]{@{}l@{}}Blind spot assistance\end{tabular} & Sensor \\
\hline
\begin{tabular}[c]{@{}l@{}}Adaptive headlights \end{tabular} & Steering input \\
\hline
\begin{tabular}[c]{@{}l@{}}ACC\end{tabular} & Laser and radar \\
\hline
\begin{tabular}[c]{@{}l@{}}Advanced driver   assistance systems \\ (ADAS)\end{tabular} & \begin{tabular}[c]{@{}l@{}}Cameras \\
\end{tabular} \\
\hline
Parking support & Radar \\
\hline
Collision mitigation & Radar \\
\hline
\end{tabular}
\end{table}

\noindent This section highlights the contribution and benefits of ITS applications, and its role in leveraging communication technologies to ensure safety, provide real-time information and optimize control system for autonomous vehicles. However, using ITS for future events, such as predicting traffic flow or traffic accidents would require using more advanced technologies, i.e., artificial intelligence applications, as detailed in the next section.

\section{AI-Driven smart mobility systems}

The existing data sources and digital platforms within an urban area generate massive amounts of traffic data in different formats and sizes. Therefore, many studies are attempting to use this bulk of data through newly developed models, including artificial intelligence models, to help in decision making and urban traffic prediction \cite{liu2018urban}. Accurate predictions contribute to improving urban transportation systems. Urban traffic prediction is the use of sophisticated machine learning models to enable wireless devices, to learn traffic pattern behavior and proactively predict traffic conditions \cite{chen2019artificial}. Furthermore, traffic modeling requires focusing on specific aspects of urban traffic forecasting such as traffic speed prediction, traffic flow prediction, and traffic accident risk prediction, which are detailed in the following sections. 
\subsubsection{Traffic speed prediction}
Traffic speed prediction is an indicator that measures future traffic conditions by calculating the average speed of traveling vehicles on each road segment, as discussed in  \cite{liu2018urban} \& \cite{jan2017deep}. Data related to traffic speed could be collected from loop detectors, cameras or GPS equipped vehicles.
The works in \cite{lv2018lc},\cite{yao2019revisiting} \& \cite{essien2019improving}  attempted to fine-tune the prediction model and improve accuracy by examining the context factors, such as temporal patterns, spatial dimension and weather conditions. 

\textit{Temporal patterns:} The work in \cite{lv2018lc} investigated the correlation between the fluctuation of speed and the temporal patterns. The study mainly focused on improving time series prediction by studying and learning speed fluctuation. It used the look-up recurrent neural network (LC-RNN) that combines a recurrent neural network (RNN), to study sequence learning, with a convolutional neural network (CNN), to capture and model complex traffic evolution in spatial. The model is based on taxi trajectory data collected through the fusion of traffic data sources with temporal patterns in a large road network in Beijing and a small road network in Shanghai. 
However, these conditions were not valid in all contexts. In fact, spatial dependence could manifest some dynamic patterns and temporal dynamics may have some perturbations. In this perspective, the work in \cite{yao2019revisiting} proposed the use of a spatio-temporal dynamic network (STDN) to learn the dynamic similarity and periodic temporal shifting.

\textit{Weather conditions:} The authors in \cite{essien2019improving} made the speed prediction more accurate by incorporating weather parameters into traffic data input. They adopted a long short-term memory neural network (LSTM-NN) as a deep learning prediction method. The experiment was conducted in Greater Manchester, in the Uk, where it investigated the impact of temperature on traffic prediction.  
Traffic data (average speed, flow, density) were collected in real-time from several sources including inductive loop sensors, by Transport for Greater Manchester (TfGM), and were used for the data fusion technique 'Data in'-'Data out' adopted in the study.\\
\noindent The traffic speed prediction was categorized into the following congestion levels  (e.g., slow, normal, and fast) \cite{liu2018urban}.

\subsubsection {Traffic flow prediction} 
Traffic flow determines the movements of a crowd by measuring the total amount of vehicles and humans that move through a defined space at a large scale during a period of time. 
Predicting traffic flow consists of analyzing data collected from infrastructures such as AFC systems.
Several researchers have attempted to estimate future crowd density by building flow predictor using a periodic convolutional recurrent network (Periodic-CRN), a predicted model that learns previous periodic representations and use them in geo-spatio-temporal domains \cite{zonoozi2018periodic}.\\
The work in \cite{cheng2018deeptransport} used CNNs and RNNs to study traffic information forecasts using road temporal and spatial information of surrounding locations to relieve traffic congestion.\\
The authors of \cite{wang2018deepstcl} focused on forecasting travel demand for transportation and traffic resources. They developed the concept of a deep spatio-temporal convolutional long short term memory (LSTM) learning scheme, which is a new deep learning model that captures both temporal and spatial characteristics.

\subsubsection{Traffic accident risk prediction}
Traffic accident risk prediction refers to the probability of an accident to occurr on a given road due to several factors such as driver behavior, weather conditions, and traffic congestion. The number of traffic accidents fatalities worldwide is becoming very alarming, and to cope with this issue, several studies have attempted to predict accidents by building models to analyze correlations between existing factors. In this respect, the work in \cite{chen2016learning} developed a traffic accident risk predictor by correlating time interval, region and human mobility density. This model allows alerting drivers of potential accidents in real-time. In \cite{lamr2018big}, the authors developed an early warning system on the possible occurrence of traffic accidents, based on a prediction model that uses a combination of traffic accident data and traffic intensity data in real-time and location. The system warns the driver when the current situation presents a high similarity with past conditions during which traffic accidents occurred.

\subsubsection{Driver behavior prediction}
In \cite{bhattacharya2018eye}, the authors designed an algorithm that predicts driver behavior, and more particularly, the likelihood to change lanes, by studying the frequency of his eyes glance. 
The experiment used a driving simulator that captures and analyzes visual behavior such as eye positions while performing secondary tasks, for instance, listening to music, talking to other passengers or using the cell phone. The study assessed the impact of secondary tasks on eye glance behavior and found out that discussing with co-passengers while driving could lead to adverse consequences.

\noindent In accordance with the aforementioned context, several initiatives have been developed to demonstrate the potential of machine learning models applied to all traffic prediction aspects (see Table \ref{Table 8}).

\begin{table}[!htb]
\caption{The table summarises the different machine learning models applied on several types of captured data (inputs) enabling to predict various aspects of urban traffic.}.
\label{Table 8}
\begin{tabular}{|l|l|l|l|}
\hline
\textbf{Category} & \textbf{Input} & \textbf{Output} & \textbf{Model} \\ \hline
\textbf{\begin{tabular}[c]{@{}l@{}}Traffic speed \\ prediction\end{tabular}} & \begin{tabular}[c]{@{}l@{}}Infrastructures\\ GPS\end{tabular} & Congestion level & \begin{tabular}[c]{@{}l@{}}LC -RNN\\ LSTM-NN\end{tabular} \\ \hline
\textbf{\begin{tabular}[c]{@{}l@{}}Traffic flow\\ prediction\end{tabular}} & AFC systems & Crowd density & \begin{tabular}[c]{@{}l@{}}Periodic-CRN\\ CNN \\ RNN\end{tabular} \\ \hline
\textbf{\begin{tabular}[c]{@{}l@{}}Travel \\ demand \\ prediction\end{tabular}} & \begin{tabular}[c]{@{}l@{}}Historical \\ observations\end{tabular} & \begin{tabular}[c]{@{}l@{}}Distribution of \\ current travel\\ demand \end{tabular} & \begin{tabular}[c]{@{}l@{}}Deep  spatio-\\ temporal  \\ convolutional  \\ long  short\\ term memory\end{tabular} \\ \hline
\textbf{\begin{tabular}[c]{@{}l@{}}Traffic risk \\ accident \\ prediction\end{tabular}} & \begin{tabular}[c]{@{}l@{}}AFC systems, \\ social\\ media data, \\ historical\\ accident \\ reports\end{tabular} & \begin{tabular}[c]{@{}l@{}}Accident risk \\ probability\end{tabular} & \begin{tabular}[c]{@{}l@{}} Correlation\\ between time \\ interval, \\positions and\\ human mobility \\risk probability\end{tabular} \\ \hline
\textbf{\begin{tabular}[c]{@{}l@{}}Driver \\ behavior \\ prediction\end{tabular}} & Driver simulator & \begin{tabular}[c]{@{}l@{}}Driver glance\\ behavior\end{tabular} & \begin{tabular}[c]{@{}l@{}}Visual \\ behavior\\ analysis \end{tabular} \\ \hline
\end{tabular}
\end{table}

\noindent On the one hand, AI-based urban traffic prediction is one of the main approaches adopted to direct the traffic flow and relieve traffic congestion. On the other hand, ITS solutions are also used to effectively improve route conditions, and thus help drivers plan their trip accordingly. Vehicle routing provides customized routing guidance by leveraging communication technologies, ITS applications, and applying AI techniques. AI combined with ITS deliver accurate predictions that help users benefit from effective route planning and navigation in their daily lives. Next, we discuss how ITS and AI contribute to enhancing vehicle routing performance, alleviating traffic congestion, and reducing accident risks. 

\section{Vehicle routing systems}

The urban transportation system could be simulated to a network of routes with different traffic capacities. It is also considered as a dynamic multi-agent system, formed by vehicles and infrastructures. A vehicle agent moves from an origin to a destination under a travel time constraint. The infrastructure agent located at a road intersection is associated with traffic lights. The first roles of the infrastructure agent is to collect intentions from vehicle agents, i.e., deadlines and destinations. For instance, vehicle agent\textit{ v1 } plans to arrive at the hospital in 10 minutes. The second role of the infrastructure agent is to provide route guidance by assigning a path correspondent to the vehicle agent, to reach the destination at the desired time. This model brings us to the definition of vehicle routing, which refers to the assignment of optimal route, defined as the shortest route with a minimum traffic delay, and lowest cost possible to the desired destination. Several route guidance approaches were proposed in \cite{cao2018multiagent}, \cite{cao2017maximizing}, \cite{cao2016multiagent}, \cite{cao2015finding}, \cite{cao2016unified}, \cite{guo2017routing}, to perform route assignment as discussed next. For instance, the authors of \cite{cao2016multiagent}, combined the probability tail model with dynamics intentions collected by infrastructure agents. The probability tail model consists of pre-computing route guidance to every single vehicle before departure. It assigns the optimal path that  maximizes the probability of arriving on time. However, other vehicles have different deadlines, which may influence others and result in increased traffic volumes due to limited road capacities. The decentralized multi-agent approach considers the intention of other vehicles, who have different deadlines, through the vehicle reporting to roadside infrastructure, then updates routes accordingly, and provides adapted route guidance to each vehicle to increase the chance of arriving on time for all vehicles. However, some scenarios illustrate the detouring routes for vehicle agents, due to the detection of road congestion. To address this routing problem, the work in \cite{cao2018multiagent} proposed to integrate both indicators, i.e., arriving on time and total travel time into the vehicle routing problem. Due to uncertainties that may occur in roads, such as bad weather, road work, traffic accident, the authors of \cite{cao2015finding}, proposed to use the cardinality minimization approach, that consists of minimizing the probability of arriving at destination later than the deadline. Addressing the problem of traffic congestion, led authors in \cite{cao2016unified} to use the pheromone model, to analyze the current and future traffic conditions, then guide other vehicles on routes. The work in \cite{cao2017maximizing} designed a Q-learning method to improve the accuracy and reliability of travel time prediction, based on Stochastic Shortest Path (SSP). The Q-learning method is based on the combination of three approaches, (1) Traffic modeling, (2) Vehicle rerouting and (3) Traffic lights control. Based on the predicted road congestion, the framework starts a rerouting strategy that consists of redirecting vehicles from congested areas to uncongested ones and exploring traffic light control.

\noindent This section presented different approaches to vehicle routing that address traffic congestion problems. The contribution of vehicle routing becomes substantial when it comes to determining the optimal charging plan for electric vehicles through a selected path. The next section presents some works related to vehicle routing for electric vehicles, and discusses some of the relevant initiatives that support e-mobility emergence in general.

\section{E-mobility systems}
The world is witnessing the slow transition from conventional fuel cars to electric vehicles (EVs) to overcome the negative impact on the environment and the economy caused by fossil-based crude oil. This global movement related to the promotion of electric mobility is supported by international strategies and national policies to build an efficient transport system by 2050 \cite{ec2011white}. New legislation has been applied to the automotive sector and led car manufacturers to produce a new generation of hybrid and electric vehicles \cite{martinez2017energy}. Many European cities have been invited to deploy networks of chargers for public electric vehicles \cite{Efthymiou2017}. However, people are still reluctant to buy electric cars due to several factors, including limited battery life and limited availability of charging points \cite{hubner2012its}.  
Governments are providing incentives and funding to encourage the adoption of electromobility. For example, the UK government has decided to fund the installation of domestic chargers to encourage the use of electric vehicles \cite{elbanhawy2014investigating}. The European Commission has funded the SmartCEM project, which uses ITS applications for electric vehicles in response to various challenges in the electric vehicle market. Other systems have been developed to save time for commuters. For example, a rapid transit system, consisting of small driverless electric vehicles traveling on a dedicated lane, has been developed to provide passengers with non-stop transit from origin to destination in urban areas \cite{fatnassi2015viability}. In addition, several countries have set up a carpooling program to promote the use of electric cars and to map the performance of the car-sharing service, including capillarity, flexibility, space, time, co-modality, rate, availability of incentives, types of vehicles, ease of access, ease of payment and reservation (See Table \ref{Table 5}) \cite{arena2014service}.\\
Car sharing programs aim to meet the specific needs of mobility. To map mobility profiles, some companies proposed several vehicle-sharing configurations to customers, which led to the characterization of the following categories of mobility profiles: commuter, shopping, neighborhood trip, business trip, tourism and nightlife \cite{arena2014service}.\\ Some countries focused on the deployment of electric vehicles in public transportation. For instance, as part of the Conference of the Parties (COP 22) held  in Morocco, in November 2016, Marrakesh became the first Moroccan and African city to launch the Rapid Transit Bus, powered by a solar power station. The project aims to connect the peripheral districts of Marrakesh by reinforcing four other fleets \cite{marrakechelectricbus}. Other Moroccan cities, such as Casablanca and Rabat, decided to upgrade their infrastructures through the construction of tram lines in a network of 110 km \cite{guidecasatransportcasa}.

\begin{table} [!h]
\caption{Examples of vehicle sharing services in Italy\cite{arena2014service}}
\label{Table 5}
\begin{tabular}{lll}
\textbf{Services} & \textbf{Characteristics} & \textbf{Incentives} \\
\hline
GuidaMi & \begin{tabular}[c]{@{}l@{}}Several parking\\   stations\end{tabular} & \begin{tabular}[c]{@{}l@{}}Service discount\\   when parking the car\\  at IKEA parking lot\end{tabular} \\
\hline
E-vai & \begin{tabular}[c]{@{}l@{}}- Railway stations  \\ - Airport parking\end{tabular} & Access to bus lanes and free parking  \\
\hline
Car2Go & No dedicated station & Access to bus lanes and free parking  \\
\hline

\end{tabular}
\end{table}

\noindent On the other hand, to allow drivers in the USA to be better informed about available charging stations, a platform, called PlugShare, has been developed with a database of more than 50,000 charging stations, which maps North American charging stations. In addition, several carpooling initiatives have been implemented to foster electric vehicle use. Table \ref{Table 6} illustrates the contribution of ITS in electric vehicles by providing relevant information to drivers \cite{yang2017ev}.

\begin{table} [!h]
\caption{ITS for electric vehicles \cite{yang2017ev}}
\label{Table 6}
\begin{tabular}{ll}
\textbf{EV market challenges} & \textbf{ITS solutions} \\
\hline
\textbf{\begin{tabular}[c]{@{}l@{}}Limited\\   charging point locations\end{tabular}} & \begin{tabular}[c]{@{}l@{}}Provide guidance to the nearest charging \\ point taking into consideration parking \\ availability and energy network capacity.\end{tabular} \\
\hline
\textbf{\begin{tabular}[c]{@{}l@{}}Limited\\   range\end{tabular}} & \begin{tabular}[c]{@{}l@{}}Calculate the range based\\ on driving style, weather \\ conditions and congestion levels.\end{tabular} \\
\hline
\textbf{\begin{tabular}[c]{@{}l@{}}Charging\\   forecasting\end{tabular}} & \begin{tabular}[c]{@{}l@{}}Schedule the charging\\   time based on driver behavior.\end{tabular}\\
\end{tabular}
\end{table}

\noindent Moreover, the routing problem for electric vehicles has been addressed in \cite{sohet2020coupled}, \cite{ma2011decentralized} and \cite{etesami2017smart}. For instance, the authors of \cite{etesami2017smart} studied the interaction of selfish EVs and developed a smart routing framework that takes into account costs related to traffic congestion, electricity price and availability, as well as waiting time at charging stations. The work in \cite{sohet2020coupled} built a model that couples EV driving decisions with charging strategies, to design incentive mechanisms for EV drivers. \\

\noindent This section surveyed the different smart mobility systems. It highlighted the important role of IoT in processing data and delivering accurate traffic information for several purposes, including traffic prediction, vehicle routing optimization, safety services, and autonomous mobility. However, the effective use of IoT requires high-quality data collection process, necessary for a reliable data analytic process. The following section presents the existing data sources in this area and underlines the importance of the degree of openness and the quality level of traffic data, considered as the ingredients for building the smart city vision.  

\section{Smart mobility from a data perspective}
Urban mobility needs to be studied from a data perspective seen as the driving force behind the concept of smart mobility. Huge amount of digital raw data is produced daily in various formats and sizes from different sources. Examples include the \textit{ Internet of Things (IoT) }that generates massive amounts of data from sensors and other smart devices. The data collected from IoT could be used to explore the central problem of urban mobility. Also, \textit{online social media} could be used to contribute to communicating data generated by millions of users; it could help understand the traveler behavior and adapt the transportation services accordingly. Many other examples showcase the substantial use and role of data in designing a sustainable and resilient city. With data proliferation, new concepts have also emerged to help bring cities closer to the smart city vision by creating products and services intended to citizens. The following section highlights one of these concepts, that might improve transportation services. 

\subsection{Open data }
This section introduces the open data concept and emphasizes the quality of data characteristics in the data value chain.
Open data is defined as data published by governments, as the official data provider, to be freely used and distributed by the public. Increasingly, governments are considering opening data to release social values in several sectors including transportation. After the publication of the open data policy in the memorandum of heads of departments and agencies \cite{obama2009memorandum}, many governments have officially announced their adhesion to the open data movement. They have made significant progress in developing sustainable mobility models, adapted to today's urban infrastructure.Transport schedules, traffic accidents and real-time transits are examples of open transport data, that enable citizens to be better informed, with improved service delivery, and a well-established communication between all stakeholders.\\

\begin{figure*}
    \centering
    \includegraphics[width=\textwidth]{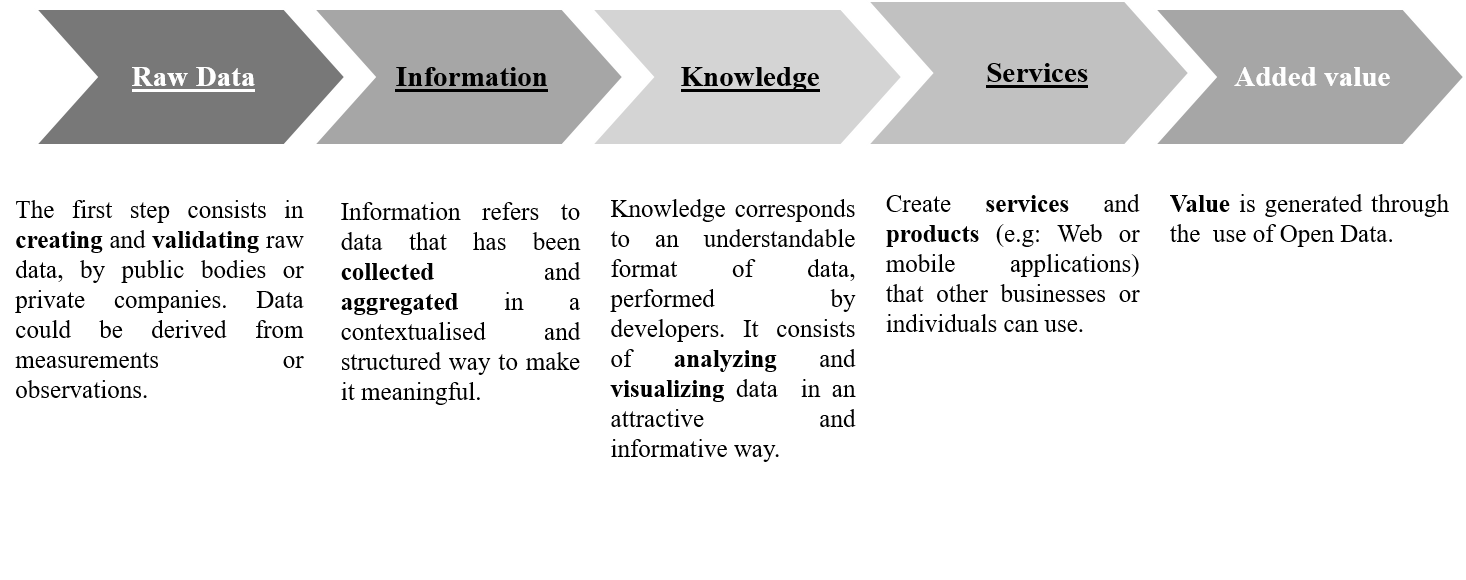}
    \caption{This figure represents a linear sequencing of open data value chain\cite{carrara2015creating}}
    \label{fig-ODVC}
\end{figure*}

\noindent Transportation services seek to create urban mobility value using open data. 
For instance, the economic value of open data in transport is estimated to be between \$720 and \$920 billion a year, and 35 hours could be saved by commuters due to schedule changes \cite{manyika2013open}. The process of deriving value from data is called an open data value chain, which indicates the necessary steps illustrating how data is processed from the creation to the added value (Figure \ref{fig-ODVC}). 

%\begin{figure*} [htb]
%\begin{center}
%\includegraphics[width=0.9\textwidth]{ODVC.pdf}
%\end{center}
%\caption{Open Data Value Chain}
%\label{fig-ODVC}
%\end{figure*}

\begin{figure}
      \includegraphics[scale=0.45]{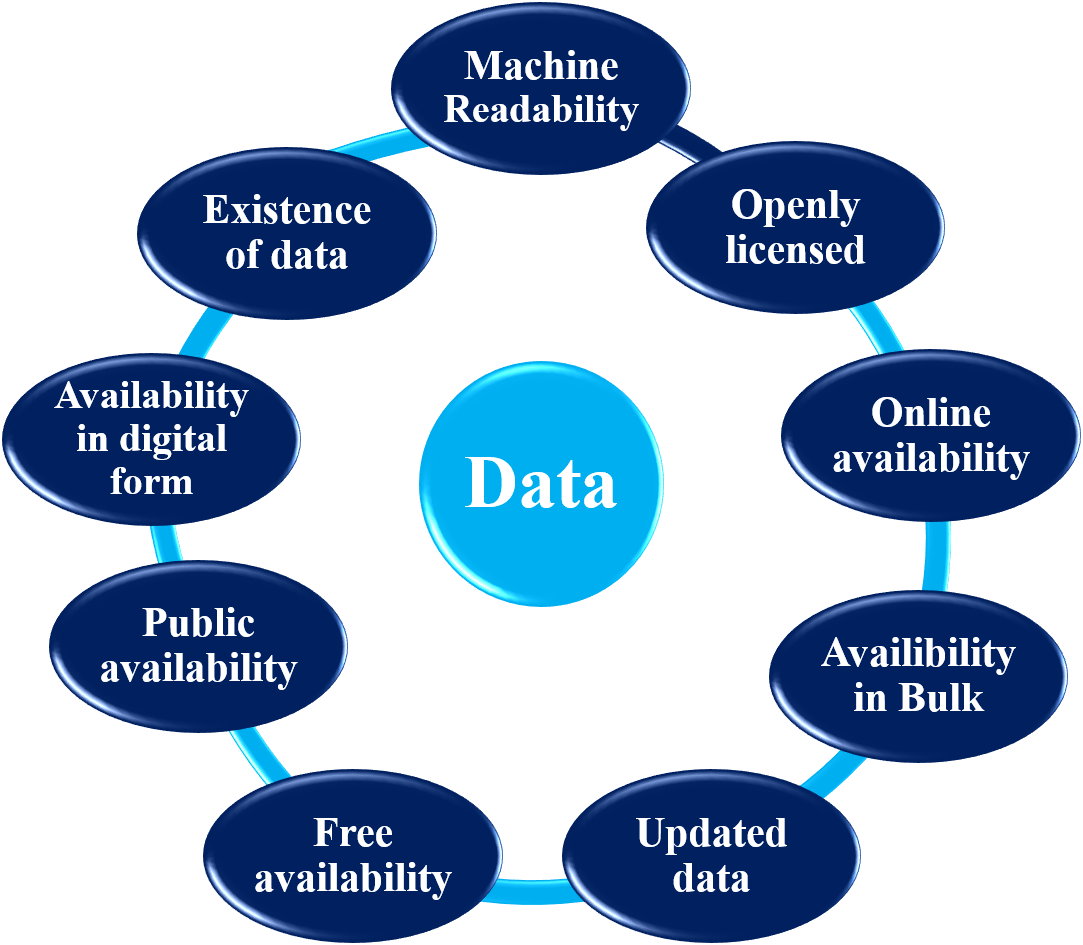}
      \caption{Open data criteria}
      \label{fig-o}
\end{figure}

\noindent The first step in the open data value chain is data creation by public bodies or private organizations. For example, London has launched its official open data portal on transportation where travel and accessibility options are displayed to allow travelers to better plan their trip \cite{TransportforLondon}. The second step consists of aggregating data from different datasets and structuring it in a meaningful format. The third step is related to the analysis and visualization of data in an understandable format. The last phase consists of designing web or smartphone applications to help users make better decisions. For instance, in New York, several apps have been developed to allow commuters to navigate the subway, save time and avoid confusion. In Sweden, real-time data on transit departures and arrivals are published, helping users to be informed and better plan their journey. The usage of open data helps create value in the transportation sector. However, the data value is mainly dependent on the degree of openness of the data that can be measured based on the following parameters (see fig. \ref{fig-o}) \cite{opendatacensus}.\\
\begin{itemize}
\item \textbf{Existence of data}: Data is published by the government or by a third party representing the government, institutions or individuals.
\end{itemize}
\begin{itemize}
\item \textbf{Public availability}: Data is available to the public in an online or printed format.
\item \textbf{Online availability}: Data is accessible via an online portal or a website.
\item \textbf{Availability in digital form}: Data is in digital format but may not be accessible online.
\item \textbf{Bulk availability}: An online platform where data are published offers the ability to download all datasets.
\item \textbf{Free availability}: No cost is associated with the access, the use or the reuse of data.
\item \textbf{Open license}: Data could be freely used, reused and distributed.
\item \textbf{Machine readability}: A computer can read and extract data from a document (unlike a scanned copy).
\item \textbf{Updated data}: Data is continuously maintained accurate.

\end{itemize}

\noindent The smart urban mobility vision relies substantially on people's decisions and behaviors in planning their trip. Many technologies and approaches have been developed to understand mobility patterns and to model urban human mobility by collecting data about travelers. Processing the collected data about social mobility helps in drawing specific conclusions. For instance, the investigation on commuters mobility behavior indicates that the factors behind modal choices for commuters are related to the crowding problem and the transport unreliability that drive most travelers to use their private car rather than public transit \cite{tyrinopoulos2013factors}.
\noindent There is a variety of techniques and methodologies to collect traffic data. Each method has a different technical characteristic that indicates the quality level of the collected data. The latter is defined by specific attributes, as detailed next:
\begin{itemize}
\item \textbf{Data reliability} is a way to ensure and maintain data consistency during the data collection process in a distributed environment \cite{igidatareliability}.
\item \textbf{The cost of data} refers to the costs associated with a data collection technique. This includes expenses related to the systems and processes required to collect the data.
\item \textbf{Data accuracy} means that the form and content of the stored data values are the correct values. The accuracy of the data is continuously related to the cost of the data collection technique. The higher the cost, the more accurate the data will be \cite{ebrarydatacollection}. 
\item \textbf{The freshness of data} is linked to the latest update of the data. It describes the time required for data collection, processing and transmission, as well as for continuous detection of malicious interpellation of the transmitted data \cite{igidatafreshness}. In the context of urban traffic data, the freshness of data indicates whether the data collected or sensed is recent \cite{igidataattribute}. The idea of the freshness of data in general, has also been recently explored in various other smart city domains such as telecommunications \cite{zhou2019joint}.
\end{itemize}
However, the collection of relevant official data on urban mobility varies according to the context of the country. For example, developed countries can afford to invest in infrastructure such as surveillance cameras to collect mobility data, unlike other countries, where the data collection process is operated at a lower cost. Some examples are illustrated below; they include the most important sources of data collection for smart mobility. 

\subsection{Data collection sources }
Data collection can be performed using a variety of means such as:

\noindent \textbf{Automatic fare collection (AFC) system}\\
The automatic fare collection system was initially used to collect travel transactions. It also serves to extract travel information data from a smart card data used by commuters, allowing the system to study their movements, estimate their travel patterns, and forecast their destinations according to the boarding and descent of passengers. The AFC system is widely implemented by several big cities, to support public transportation service planning, and improve service quality. For instance, the work in \cite{wang2016travel} showcases the effective use of the AFC system in inferring trip chain information of commuters of Beijing, based on smart card travel behavior data. However, many studies used the AFC system to infer the destination based only on the temporal and location information extracted from smart card data. To enhance the accuracy of destination inference, the work described in \cite{jung2017deep} considered incorporating land-use characteristics into travel data of commuters, by using a deep learning model.\\

\noindent \textbf{Automatic toll collection system}\\
The automatic toll collection system represents a fully automated road pricing system that collects toll from vehicles and gathers traffic data in near real-time at the highway. It is used as a source of data for several purposes, for example, in \cite{wan2016predictability}, the collected data are used to predict vehicle mobility and traffic flows along the expressways of Beijing. In Taiwan, the authors of \cite{fan2018using} collected data from the automatic toll collection system to predict highway travel time and provide drivers with estimated and adjusted travel time information. In Austria, the experience in \cite{schneider2009traveller} consisted of collecting toll data from heavy vehicles in highway toll stations to provide near real-time information on the traffic situation on all Austrian highways (4,000 km). The automatic toll collection system is also useful in addressing social problems. For instance, the work in \cite{christopher2019smart} showcases its use in alleviating traffic congestion by automating the process using an android application and Radio frequency identification (RFID) smart cards, which contributes to decreasing fuel utilization, air contamination and wastage of time. AFC systems and automatic toll stations are both considered to be secure, recent, accurate and reliable data sources.\\

\noindent\textbf{Global positioning system (GPS)}\\
A GPS system captures data on real-time positions of vehicles. The collected GPS data are then used in a wide variety of applications, including real-time tracking and alerting system, forecasting travel time information, identifying congested locations, reporting speed and position of each vehicle, and providing online map navigation services. Practically, the work in \cite{ghosh2017intelligent} illustrates the effective use of GPS data in providing better public transportation services, and particularly buses, allowing users to access real-time information through a web or mobile application about the bus route and its position in real-time. In that way, users can plan their journey and would be able to rapidly and easily reach their destinations. In another context, GPS data could be used to study social community behavior, as investigated by authors in \cite{castro2013taxi}, where the study is based on traces of taxis equipped with GPS. Examples of taxi GPS traces include: pick-up and drop-off location, route choices, strategy to find a new client, type of connections between several areas of the city, and identification of cities hot spots. They are considered all crucial information that helps in planning the city's infrastructures.
However, some GPS data problems may arise. For instance, due to the lack of GPS signals in some locations, GPS data is received with a lapse of several minutes. GPS data are then missing or erroneous.\\

\noindent\textbf{Smartphones}\\
The increasing use of mobile technologies and the multi-functionality of smartphones have led to the production of several categories of data, such as telecommunication data, crowdsensed data, and driver behavior data, that are all mainly used for urban analysis. A smartphone is considered as a mobile sensor that leaves digital traces and allows the detection and collection of dynamic environmental data (image, video, geographic position, temperature). The collected data is then converted into traffic flow information and is used for urban sensing applications. The work in \cite{calabrese2015urban}, provides an overview of the value of telecommunication data in understanding mobility patterns, studying the exploitation of urban spaces, identifying types of activities and distribution of population across the city, as well as analyzing travel demand during particular events. Practically, the experience conducted in \cite{Tosi2017} showcases the concept of computing mobility patterns from cellular data in Milan. Data is collected from the cellular network of the Vodafone Italy Telco operator to observe people's behavior and then inform users of traffic information.\\
Crowdsensing data is an alternative approach to data collection. It consists of exploiting sensing capabilities offered by smartphones to produce aggregate information about a given situation. For instance, collecting geospatial data from the crowd is significant to create route planning and navigation services. Sensed data could be video clips, photos, measurements, or texts \cite{liu2016data}.
Mobile crowdsensing has many advantages, such as cost-effectiveness and the provision of very recent sensed data. However, the crowdsencing approach presents many risks that could compromise the accuracy, reliability and quality of crowdsensed data, as well as the possibility of providing false information \cite{stojanovic2016mobile}.\\
As for data about driver behavior, the authors of \cite{meseguer2017drivingstyles} used smartphones combined with the vehicle electronic control unit (ECU) through the on-board diagnostics (OBD-II) Bluetooth interface to collect data about driving characteristics including speed, acceleration, throttle position, and vehicle's geographic position. The data analysis reveals the degree of aggressiveness of drivers and characterizes the different existing driving styles. Based on data processing results, the system monitors and helps drivers in real-time to adopt the right behavior to optimize fuel consumption and achieve fuel savings of 15 to 20\%.
Smartphones are inexpensive devices for data collection, in comparison with other materials that call for investment. However, smartphones may not deliver accurate information due to several reasons such as network coverage issues or problems in sensing capabilities.\\  

\noindent\textbf{Social media}\\
Social media represents a different type of data source that relies explicitly on users' daily information, which can take the form of tags, tweets, likes, photos, perceptions about a given urban space, and places visited, or mobility needs. Social media data have georeferenced characteristics and are used to provide a better understanding of mobility and social behavior within urban areas. Several studies have been conducted to investigate mobility behavior based on social media data. This approach is illustrated by the work in  \cite{assem2017rcmc} that studied crowd mobility patterns at the level of New York City from a social media perspective using location-based social network data (LBSN). LBSN  refers to the activity data or check-in data shared by users through social media. For example, when a user shares a tweet, the following variables are recorded: user ID, tweet ID, time, shared content, followers, location. Social media data enables one to study people's mobility at a larger scale over a time period, and understand motivations behind the crowd shifting from one place to another. Data results will enable the city stakeholders to allocate city resources efficiently.\\

\noindent \textbf{Camera surveillance}\\ 
Governments and local authorities use cameras for intelligent detection and traffic surveillance. Cameras are placed on busy roads, intersections, tunnels, or highways to store, record video, or capture images to detect incidents or collect traffic data \cite{sharma2017sustainable}.
The many benefits of camera surveillance include capturing reliable live data. However, camera surveillance is expensive in terms of hardware infrastructure, which consists of the monitoring process, capture infrastructure, data transport and storage.\\

\noindent One of the promising approaches to ensure high-quality data processing is to combine multiple data sources. For example, to have more accurate travel time information, the authors of \cite{zhang2018real} implemented a multi-view model technique that includes mobile phone data and transport data (taxis, buses and subways). With the same approach, the work in \cite{long2019predicting} has collected various types of data from multiple sources (toll data, weather data, traffic accident data) to predict travel time on the freeway of China. Moreover, several new systems and models have been developed to extract and use real-time traffic data to obtain relevant information on current traffic conditions. For instance, the authors in \cite{zheng2014urban} and \cite {wang2017computing}, built a data integration system, that consists of combining several real-time urban data collected from sensors and social media to obtain a complete and accurate estimate of traffic congestion in real-time. The system helps operators to understand the root cause of urban problems by mining environmental, energy, economic, transportation, social security, and urban planning data. Furthermore, the work in \cite{manolopoulos2013smartphone} sought to implement a new Traffic information system (TIS), based on smartphones and is used to estimate traffic conditions while meeting safety and privacy requirements. Unlike other costly TIS based on road infrastructures, the developed system is dedicated to collect data on vehicle speed with broader coverage. TIS allows drivers to stay informed about road traffic conditions. For instance, the driver receives the following information:\\ 
- Smooth traffic when the speed is greater than 25 km/h;\\
- Medium traffic when the speed is between 14 and 25 km/h;\\
- Congested when the speed is less than 14 km/h.  \\

\noindent The previous sections outlined the most relevant smart mobility systems using various technologies such as ITS, AI, routing vehicles, and have highlighted the importance of smart data in building such systems. However, some countries still face the problem of technology transfer in smart mobility due to several factors. Some of the technology differences across countries are discussed in the next section.

\section{Smart mobility in Developed Countries Vs. Emerging Countries}

Developing a smart mobility strategy may differ from one city to another. The strategy depends on the political-economic and socio-cultural context of each city, which includes several factors such as quality infrastructures, transparency degree, level of citizen engagement and citizen readiness, environmental impact and a number of social partnership initiatives.\\

\noindent\textbf{Infrastructure}\\
In developed countries, cities are qualified to be mature in terms of ICT infrastructures, which give them a leading edge over emerging cities that are still at the stage of upgrading and modernizing transport system infrastructures. The development of the transportation sector in developing countries is hampered by the limited capacity of cities in terms of human, technical and institutional resources. It's also hindered by the lack of commitment to promoting public services related to transport systems and infrastructure \cite{morocco2014programforresultstransport}.
The poor quality of road infrastructure and the aggressive behavior of drivers are at the root of many traffic problems. Concrete examples show that the current urban infrastructure is not adapted to buses (no corridor dedicated to buses). As a result, buses, taxis and passenger cars share the same roads, which aggravates traffic congestion. The same problem applies to emergency vehicles, which need a dedicated infrastructure to help them play their role effectively in critical situations. On the other hand, in some areas of a city, where several lanes intersect, the traffic lights appear unbalanced, which creates significant traffic in each lane and irritates drivers who adopt aggressive behavior. \\

\noindent\textbf{Transparency}\\
In developed countries, open data transportation becomes a regular tool used to  manage traffic and allow travelers to better plan their trips. Past experiences show the effective contribution of open data in leveraging the economy in the transportation sector. In contrast, in emerging countries, the urban transport system suffers from several problems related to the lack of transparency of passenger information in real-time, and the lack of reinvestment in public transport due to low financial viability \cite{morocco2015urbantransportprogram}. 
\noindent Recently, open data has caught the attention of governments of emerging countries, where several data platforms with many datasets were created. However, the mobility data still suffer from scarcity, and the handful of published data are sometimes obsolete. It is essential to have the political will and to encourage all economic actors to adhere to the open data policy. This helps improve the efficiency of public transport services.\\

\noindent\textbf{Citizen Engagement}\\
In developed countries, city governments rely significantly on citizen feedback to choose the key pillars of smart city plans. It consists of adopting a collaborative approach between elected representatives, civil society organizations and citizens, to co-create a smart city action plan. Concretely, in Vancouver, citizens were highly involved in the process of diagnosing the needs to help build a smart city. They were asked to provide their priorities and proposals to make sure that the city plan is aligned with their expectations. The main priority highlighted by citizens was related to smart mobility. In the same context, the city of Reykjavik, has crowdsourced ideas from citizens through an online platform, and has invited them to suggest a budget allocation for each idea. This process had helped many thousand citizens to feel the weight of their influence in building their environment \cite{top50}. 

\noindent In emerging countries, participatory democratic tools have been implemented (e.g., petitions, public consultation, online crowdsourcing platforms) to help citizens guide city governments in developing smart mobility strategy. For instance, Jaipur city \cite{top50} has established a consultation campaign to collect citizens' suggestions about smart and sustainable solutions for infrastructures. \\

\noindent\textbf{Social Partnership}\\
Numerous initiatives have experienced the social partnership model as a key element to enable building a smart city transport system. For instance, in the Netherlands, the city of Amsterdam, has strengthened its expertise from an online academy, where smart city tools and methodologies are available, and are used by public officials. In the United States, the city of Columbus, Ohio, has decided to partner with private sector institutions, who are in charge of funding the large smart city projects that include particularly smart transportation solutions adapted to the needs of residents. In addition, new models of smart mobility have emerged in Helsinki and Tallinn, referred to as joint smart mobility projects, where city governments joined forces to solve everyday challenges related to traffic congestion \cite{top50}.\\
In emerging countries, several initiatives were launched to promote the social partnership concept to reshape cities. One exemplary case is the ecocitizen world map project, implemented in Colombia, Morocco, and Egypt. This project is based on the concept of participatory action research, which aims to involve all stakeholders (citizens, civil society, the public sector, the private sector and the academic sector) in the areas of the environment and sustainable development. The project relies on the use of online crowd mapping, geographic information system (GIS) and social media applications that actively engage citizens in learning, sharing and applying sustainability principles and practices \cite{nesh2015citizen}. \\

\noindent\textbf{Citizen Readiness}\\
In developed countries, governments and private sector focus on the needs of all type of vulnerable users, including people with disabilities, users with reduced mobility, elderly people, expectant mothers. Cities in developed countries often invest in the well-being and quality of life of its citizens, including those with specific needs, to move around easily. Mobility facilities are implemented by removing physical barriers and enabling access to public transportation services through properly designed applications and transportation solutions adapted to vulnerable users' situation. 
In brief, cities in developed countries are offering an appropriate and comfortable environment for its residents, which enables people to use the amenities for every mobility need, regardless of their situation.
In contrast, in emerging countries, cities are still in the process of eliminating the main barriers that hinder the inclusiveness and accessibility to public transportation services. These barriers include the lack of urban accessibility for vulnerable users, a security issue that may discourage car-sharing initiatives, and digital illiteracy, which may lead to a negative impact on travel planning. 
Despite the profusion of much available information in hand, city officials are facing an unbalanced supply-demand situation in transportation services, that are manifested by troubles in using mobile applications, and resistance to change. Meanwhile, many digital transport services, including Uber, Careem, and i-taxi, have been set up in emerging cities. However, their use remains very limited when compared to developed countries.\\

\noindent\textbf{Environment}\\
The urban expansion has caused adverse consequences, in particular atmospheric concentrations of greenhouses gases (about 60\% of gas emissions are coming from urban transport \cite{toledo2018urban}).
In developed countries, cities are typically empowered to address those environmental issues. Governments are setting up national charters, and roadmaps that focus on transportation decarbonization to ensure sustainability in mobility.
For example, residents are encouraged to use public transportation instead of private cars, through an established incentive system.
However, in emerging countries, environmental issues are still not a priority, which has prompted some non-governmental organizations (NGOs) to take the lead on environmental issues. For instance, several NGOs have joined efforts to organize massive awareness campaigns, targeting children, youth, and communicate on their activities in protecting the environment through media programs or school training. NGOs can potentially pressure local governments through advocacy campaigns to cooperate and be engaged in preserving the environment.\\

\noindent In addition to the numerous efforts undertaken to face urban mobility challenges, developed countries can still upgrade their public transit by moving to the next level of aeronautic industry. This applies to the concept of hyperloop, a high-speed rail (1200 km per hour), that consists of propelling pods through vacuum tubes using magnetic levitation. It represents a kind of a proposed mode for passengers that offers cost and time savings \cite{ross2015hyperloop}. The hyperloop concept is being tested in several countries including the USA, France and the UAE.\\
In another context, air taxi services provided by Uber Copter are now open to the public in the United States and the United Arab Emirates to avoid traffic congestion and offer a better service for users.\\
While developed countries are upgrading the whole mobility infrastructures, emerging countries are still dealing with the unbalanced transportation problem. To cope with these developments, developing countries should promote joint collaboration from all sectors including, public, private and academic sectors, as well as citizens who should be at the core of the ecosystem. 

\section{Open problems}

In this section, the discussion points out to some open problems that hinder the implementation of smart mobility.

\subsection{Open problems related to ITS}
\noindent \textbf{Communication latency} \\
Although 5G cellular systems can provide low latency (1 millisecond), there is still some network coverage problems due to the geographic disparity of internet accessibility. These problems typically exist in rural areas. In addition, cars using different wireless telecommunications services may not react instantly in the same way. According to some analysts, ITS applications may not be able to provide a quick and effective response to coordinate real-time operations during emergencies, especially during natural disasters or terrorist attacks. Because in such situations, ITSs are expected to process data streams in a short time and provide accurate, driver-specific information to help manage the crisis period \cite{behruz2013challenges}. One potential solution here is through the use of flying drones equipped with wireless communication capabilities that complement ground infrastructure \cite{mozaffari2019tutorial}, \cite{chen2017caching} and \cite{mozaffari2017mobile}.\\

\noindent \textbf{Cyberattacks against ITS.}\\
Despite the many benefits of ITS to improve urban mobility, some risks may arise. For instance, collected data about users, infrastructures and vehicles, as well as all smart city clouds, are all subject to security threats. These attacks can target:
1) Road infrastructure, for example, hacking the traffic message displayed on traffic signs,
2) Wireless communication, by hacking information shared via wireless communication systems (V2X),
3) Networks, for example by compromising automated payment systems and attacking surveillance cameras, and 4) Connected or autonomous vehicles and their data which can be compromised through data injection attacks \cite{ferdowsi2019cyber}.
The following examples illustrate real ITS attacks in several cities around the world \cite{huqcyberattacks}:
\begin{itemize}
	\item 
In Dallas, a sign was placed on a highway to warn users of upcoming constructions that could cause traffic jams. A hacker guessed the login information and changed the sign into a joke message saying, "Drive Crazy Yall".
	\item 
In San Francisco, a crypto-ransomware attacked all ATMs in the subway stations and took them out of service. Passengers could not pay for the services and were offered free transportation by the municipal transport agency.
	\item 
In Washington DC, a ransomware attacked surveillance cameras, preventing the local police from recording videos for four days.
	\item 
In Dusseldorf, Germany, the public transport company failed to modernize its system due to the hacking of public transport networks, which resulted in the posting of incorrect information via digital signage in front of buses and trains. This situation resulted in significant delays and cancellation of bus and train services.
\item A hacking experiment was conducted on a Jeep Cherokee, an autonomous connected vehicle, by attacking the control system and taking remote control of acceleration and braking systems. In this example, the cyber attack can also negatively impact the flow of other vehicles\cite{ferdowsi2019cyber}.
	\item 
University researchers conducted an experiment on the behavior of autonomous cars by reading signs with stickers. For example, an autonomous vehicle continues to drive when a stop is planned. The vehicle was misled due to the misinterpretation of the message on the panel because the stickers deceived the vision system of the vehicle.
\end{itemize}

\noindent Attacks on the ITS ecosystem could have negative consequences and affect public safety, create confusion for drivers, increase congestion and cause accidents.
\\

\noindent \textbf{Conflict information. }
The presence of multiple security systems can generate conflicting information and uncoordinated messages about potential hazards. This can place an additional burden on drivers and lead to confusion and mistrust of warnings. Managing data from multiple sources simultaneously, such as those provided by a vehicle's embedded device and road infrastructure, remains a challenge for vehicular networking technology. The following example illustrates the concrete situation regarding conflicting information. The cooperative intersection collision avoidance system-signalized left turn assist (CICAS-SLTA) informs the driver that the road is free to turn left. At the same time, the red-light violator indication) (RLVI) warns of dangerous conditions due to the presence of red light on the right. Conflicts between connected vehicle safety applications can put drivers at risk \cite{richard2015multiple}. 

\subsection{Open problems related to autonomous vehicles}

\noindent \textbf {Social, legal, and economic challenges of AVs}\\
\noindent The emergence of AVs in the automotive market is expected to change the mobility ecosystem that will comprise new digital socio economic and legal changes. It would require redefining a new mindset and ethical values as well as a new legal framework and business models. Policymakers are left behind by rapid technological change and are afraid of not being able to control mobility in the city. Instead of planning the arrival of disruptive technologies, they create obstacles to better manage mobility in the city. According to Harvard University \cite{martylegal}, policymakers are wary of potential problems such as the loss of public revenues, the decrease in the number of traffic violations, parking meters, fines for traffic offenses, fuel consumption, insurance, road tax and driver's licenses, which are not considered to be profitable for some economies. On the social side, many people are at risk of job loss due to AVs. For instance taxi drivers fear losing their jobs with the emergence of driverless taxis. The consequences of such problems lead to the adoption of strict laws on autonomous vehicles and the shutdown of business services related to autonomous vehicles. Many studies and attempts have been carried out to investigate the applicable regulations related to AVs; as of today, no distinct legal system has been implemented. In addition, policymakers around the world are still discussing new legislation on the liability in the event of an accident involving autonomous vehicles. Therefore, to be fully aligned with the rapid changes of the robotic era, there is a need for lawyers and scientific experts to work collaboratively in creating an effective legal and ethical framework.\\ 

\noindent \textbf {Shortcomings of AVs sensors.}\\
AVs are still dealing with many technological challenges as outlined in Table \ref{table_example}.\\

\begin{table}[]  
\caption{Shortcoming of sensors in AVs}
\label{table_example}
\begin{tabular}{lll}
\textbf{\begin{tabular}[c]{@{}l@{}}Monitoring\\   Technologies\end{tabular}} & \textbf{Range} &  \\
\hline
Lidar & \begin{tabular}[c]{@{}l@{}}-Short ranges or\\   distances \\   -Laser beams get\\   confused when they get\\   absorbed by snowflakes \\ and water droplets\\   -Reflectivity issues\\   can be high but is on downward trend\\   -Easily get covered in snow, \\ affecting their ability to see.\end{tabular} &  \\
\hline
Camera-based system & \begin{tabular}[c]{@{}l@{}}-Functional for long\\   ranges and covers long distances.\\   -Low Image processing\\   capabilities in comparison \\ with human brain\end{tabular} &  \\
\hline
Radar & \begin{tabular}[c]{@{}l@{}}-Transmission and reception of\\   radio waves\\   -Reflectivity issues\\   -Only able to detect\\   metallic objects \\ (Pedestrians remain invisible)\end{tabular} &  \\
\hline
Ultrasonic & \begin{tabular}[c]{@{}l@{}}-High-frequency acoustic\\   waves\\   -Accurate short range\\   data (1-10m)\\   -Low cost\\   -Back up warning\\   system and parking \\ assistance systems\end{tabular} &  \\
\hline
Infrared sensors & \begin{tabular}[c]{@{}l@{}}-Lane marking\\   detection\\   -Coverage range limited\\   -Used to detect pedestrians\\   and bicycles, particularly at night\end{tabular} &  \\
\hline
\begin{tabular}[c]{@{}l@{}}Geographic positioning\\   systems(GPS)\end{tabular} & \begin{tabular}[c]{@{}l@{}}Tall building represent \\ sky obstacles and GPS\\  cannot provide accurate\\  location information to users. \\ Asadi et al\end{tabular} &  \\
\hline
\begin{tabular}[c]{@{}l@{}}Inertial navigation \\ systems  (INS)\end{tabular} & \begin{tabular}[c]{@{}l@{}}Calculate the\\   position, velocity,\\   and orientation \\ (i.e., direction and speed\end{tabular} & 
\end{tabular}
\end{table}

\noindent \textbf{Mixed fleet (AVs and non-AVs)}\\
There is a risk of an accident when a human-powered vehicle and an autonomous vehicle move in the same space. For example, when a human driver breaks the rules of the road, especially at intersections (e.g., at red lights), an autonomous vehicle brakes instantly but ultimately hits a driver-operated car. One option to avoid the accident is to turn around and head to the side of the road, but unfortunately, the impact on road infrastructure or pedestrians is also unavoidable. This risk has led some countries to consider the creation of roads and highways dedicated to autonomous vehicles. Theoretical analysis of the operation of mixed fleet traffic is also of interest to understand how human behavior can interfere with artificial intelligence.\\

\subsection{Open problems related to Electric vehicles}

\noindent \textbf{Battery life of EVs}\\ 
Lithium-ion batteries in electric vehicles have important environmental consequences. For example, the early stage of battery creation generates a significant amount of pollution from fossil fuel extraction. The same problem occurs when the battery reaches its end of life.\\

\noindent \textbf{EV charging}\\
In addition to distance anxiety, people face congestion problems at public charging stations due to the lack of enough charging points. Queues form when charging stations are fully occupied.
Another social problem is that some drivers plug in the charger and go shopping nearby for the next 3 hours. Some early work \cite{etesami2017smart} has already explored how queues related to EV charging can build up in presence of human drivers.EV charging also poses many challenges on the power system itself that must be addressed for a large-scale deployment of EVs.

\section{Conclusion}
In this paper, we have described some of the relevant initiatives related to the implementation of the concept of smart mobility in sprawling cities. In a context where the landscape of transportation is changing rapidly, and where people are very demanding, advanced technologies become a necessity. We have provided practical examples of ITS ensuring driver safety and helping traffic management through real-time services, and discussed the issues of ITS stability and latency as well as its combination with AI. This combination helps provide accurate predictions,especially in the context of autonomous mobility.  We have also described the vehicle routing problem and how to provide real-time guidance using the collected data. Besides, we have highlighted the importance of open data in improving mobility services, which can remedy transportation inefficiencies, as it can increase productivity and save time for individuals. In this regard, the transition to smart mobility would require the participation of communities in local governance. Citizens must be at the heart of local decisions and local officials must pay more attention to citizens' requests for mobility. 
Further studies should examine the complexity of implementing smart mobility initiatives in small cities compared to large cities. Small smart cities have many advantages because in local communities, people are more connected and have developed a collective identity that allows them to align themselves according to their mobility needs. 

\bibliographystyle{ieeetr}
\bibliography{bibliography.bib}

\begin{thebibliography}{100}

\bibitem{cytron2010role}
N.~Cytron {\em et~al.}, ``The role of transportation planning and policy in
  shaping communities,'' {\em Community Investments}, vol.~22, no.~2,
  pp.~2--44, 2010.

\bibitem{ECGDP2018}
T.~E.~C. science and knowledge service, ``Transport sector economic analysis.''
  \url{https://ec.europa.eu/jrc/en/research-topic/transport-sector-economic-analysis},
  2018.

\bibitem{GDPUSA}
T.~Economics, ``United states gdp from transportation and warehousing.''
  \url{https://tradingeconomics.com/united-states/gdp-from-transport}, 2019.

\bibitem{UNpopulationprojection}
UN, ``68\% of the world population projected to live in urban areas by 2050.''
  \url{https://www.un.org/development/desa/en/news/population/2018-revision-of-world-urbanization-prospects.html},
  2018.

\bibitem{economy2014better}
N.~C. Economy, ``Better growth, better climate,'' {\em The New Climate Economy
  Report, The Global Commission on the Economy and Climate}, 2014.

\bibitem{toledo2018urban}
A.~L.~L. Toledo, E.~L. La~Rovere, {\em et~al.}, ``Urban mobility and greenhouse
  gas emissions: Status, public policies, and scenarios in a developing economy
  city, natal, brazil,'' {\em Sustainability}, vol.~10, no.~11, pp.~1--24,
  2018.

\bibitem{cookson2018inrix}
G.~Cookson, ``Inrix global traffic scorecard,'' {\em INRIX Research}, 2018.

\bibitem{ECmobility}
T.~E. Commission, ``Urban mobility.''
  \url{https://ec.europa.eu/transport/themes/urban/urban_mobility_en}.

\bibitem{worldhealthorganisation2015}
W.~H. Organisation, ``Global status report on road safety 2015,'' ~, World
  Health Organisation, 2015.

\bibitem{SDGs}
U.~N.~E. Programme, ``Why does transport matter?.''
  \url{https://www.unenvironment.org/explore-topics/transport/why-does-transport-matter}.

\bibitem{ec2011white}
E.~E. Commission {\em et~al.}, ``White paper roadmap to a single european
  transport area towards a competitive and resource efficient transport
  system,'' {\em COM (2011)}, vol.~144, 2011.

\bibitem{benevolo2016smart}
C.~Benevolo, R.~P. Dameri, and B.~D'Auria, ``Smart mobility in smart city,'' in
  {\em Empowering Organizations}, pp.~13--28, Springer, 2016.

\bibitem{faria2017smart}
R.~Faria, L.~Brito, K.~Baras, and J.~Silva, ``Smart mobility: A survey,'' in
  {\em Proc. of the International Conference on Internet of Things for the
  Global Community (IoTGC)}, pp.~1--8, IEEE, 2017.

\bibitem{al2019smart}
F.~Al-Turjman and A.~Malekloo, ``Smart parking in iot-enabled cities: A
  survey,'' {\em Sustainable Cities and Society}, p.~101608, 2019.

\bibitem{bayram2014survey}
I.~S. Bayram and I.~Papapanagiotou, ``A survey on communication technologies
  and requirements for internet of electric vehicles,'' {\em EURASIP Journal on
  Wireless Communications and Networking}, vol.~2014, no.~1, p.~223, 2014.

\bibitem{isosmartcity2014}
I.~J. 1, ``Smart cities prelimineray report,'' {\em ISO}, 2014.

\bibitem{bravo2016smart}
Y.~Bravo, J.~Ferrer, G.~Luque, and E.~Alba, ``Smart mobility by optimizing the
  traffic lights: A new tool for traffic control centers,'' in {\em Proc. of
  the International Conference on Smart Cities}, pp.~147--156, Springer, 2016.

\bibitem{al2015applications}
E.~Al~Nuaimi, H.~Al~Neyadi, N.~Mohamed, and J.~Al-Jaroodi, ``Applications of
  big data to smart cities,'' {\em Journal of Internet Services and
  Applications}, vol.~6, no.~1, p.~25, 2015.

\bibitem{guo2015integration}
W.~Guo, Y.~Zhang, and L.~Li, ``The integration of cps, cpss, and its: A focus
  on data,'' {\em Tsinghua Science and Technology}, vol.~20, no.~4,
  pp.~327--335, 2015.

\bibitem{cheng2017fuzzy}
J.~Cheng, W.~Wu, J.~Cao, and K.~Li, ``Fuzzy group-based intersection control
  via vehicular networks for smart transportations,'' {\em IEEE Transactions on
  Industrial Informatics}, vol.~13, no.~2, pp.~751--758, 2017.

\bibitem{arena2019overview}
F.~Arena and G.~Pau, ``An overview of vehicular communications,'' {\em Future
  Internet}, vol.~11, no.~2, p.~27, 2019.

\bibitem{zeng2019jointplatoon}
T.~Zeng, O.~Semiari, W.~Saad, and M.~Bennis, ``Joint communication and control
  for wireless autonomous vehicular platoon systems,'' {\em IEEE Transactions
  on Communications}, vol.~67, no.~11, pp.~7907--7922, 2019.

\bibitem{saad2019vision}
W.~Saad, M.~Bennis, and M.~Chen, ``A vision of {6G} wireless systems:
  Applications, trends, technologies, and open research problems,'' {\em IEEE
  network}, 2019.

\bibitem{ferdowsi2019deep}
A.~Ferdowsi, U.~Challita, and W.~Saad, ``Deep learning for reliable mobile edge
  analytics in intelligent transportation systems: An overview,'' {\em ieee
  vehicular technology magazine}, vol.~14, no.~1, pp.~62--70, 2019.

\bibitem{noble2016influence}
A.~M. Noble, T.~A. Dingus, and Z.~R. Doerzaph, ``Influence of in-vehicle
  adaptive stop display on driving behavior and safety,'' {\em IEEE
  Transactions on Intelligent Transportation Systems}, vol.~17, no.~10,
  pp.~2767--2776, 2016.

\bibitem{hubner2012its}
Y.~H{\"u}bner, P.~Blythe, G.~Hill, M.~Neaimeh, and C.~Higgins, ``Its for
  electric vehicles-an electromobility roadmap,'' in {\em Proc. of the IET and
  ITS Conference on Road Transport Information and Control (RTIC 2012)},
  pp.~1--5, IET, 2012.

\bibitem{sumalee2018smarter}
A.~Sumalee and H.~W. Ho, ``Smarter and more connected: future intelligent
  transportation system,'' {\em IATSS Research}, vol.~42, no.~2, pp.~67--71,
  2018.

\bibitem{hamza2008dynamic}
G.~L. Hamza-Lup, K.~A. Hua, M.~Le, and R.~Peng, ``Dynamic plan generation and
  real-time management techniques for traffic evacuation,'' {\em IEEE
  Transactions on Intelligent Transportation Systems}, vol.~9, no.~4,
  pp.~615--624, 2008.

\bibitem{song2017deepmob}
X.~Song, R.~Shibasaki, N.~J. Yuan, X.~Xie, T.~Li, and R.~Adachi, ``Deepmob:
  learning deep knowledge of human emergency behavior and mobility from big and
  heterogeneous data,'' {\em ACM Transactions on Information Systems (TOIS)},
  vol.~35, no.~4, p.~41, 2017.

\bibitem{regan2001intelligent}
M.~A. Regan, J.~Oxley, S.~Godley, and C.~Tingvall, {\em Intelligent transport
  systems: Safety and human factors issues}.
\newblock No.~01/01, 2001.

\bibitem{dennyv2isafety}
R.~K. Denny~Stephens, Jeremy~Schroeder, ``Vehicle-to-infrastructure (v2i)
  safety applications performance requirements, vol. 3, red light violation
  warning,'' tech. rep., 2015.

\bibitem{stevens2012benefits}
A.~Stevens and J.~Hopkin, ``Benefits and deployment opportunities for
  vehicle/roadside cooperative its,'' 2012.

\bibitem{richard2015multiple}
C.~M. Richard, J.~F. Morgan, L.~P. Bacon, J.~S. Graving, G.~Divekar, and M.~G.
  Lichty, ``Multiple sources of safety information from v2v and v2i:
  Redundancy, decision making, and trust-safety message design report,'' tech.
  rep., 2015.

\bibitem{valldorf2008advanced}
J.~Valldorf and W.~Gessner, {\em Advanced Microsystems for Automotive
  Applications 2008}.
\newblock Springer, 2008.

\bibitem{itssafetysolutions}
R.~US~Department~of transportation and innovative~technology administration,
  ``Intelligent transportation systems safety solutions, preventing crashes and
  saving lives,''

\bibitem{hoong2012road}
P.~K. Hoong, I.~K. Tan, O.~K. Chien, and C.-Y. Ting, ``Road traffic prediction
  using bayesian networks,'' 2012.

\bibitem{li2017city}
W.~Li, D.~Wolinski, and M.~C. Lin, ``City-scale traffic animation using
  statistical learning and metamodel-based optimization,'' {\em ACM
  Transactions on Graphics (TOG)}, vol.~36, no.~6, p.~200, 2017.

\bibitem{lin2012service}
S.-Y. Lin, K.-M. Chao, and C.-C. Lo, ``Service-oriented dynamic data driven
  application systems to urban traffic management in resource-bounded
  environment,'' {\em ACM SIGAPP Applied Computing Review}, vol.~12, no.~1,
  pp.~35--49, 2012.

\bibitem{clausen2017assessment}
P.~Clausen, P.-Y. Gilli{\'e}ron, H.~Perakis, V.~Gikas, and I.~Spyropoulou,
  ``Assessment of positioning accuracy of vehicle trajectories for different
  road applications,'' {\em IET Intelligent Transport Systems}, vol.~11, no.~3,
  pp.~113--125, 2017.

\bibitem{drawil2013gps}
N.~M. Drawil, H.~M. Amar, and O.~A. Basir, ``Gps localization accuracy
  classification: A context-based approach,'' {\em IEEE Transactions on
  Intelligent Transportation Systems}, vol.~14, no.~1, pp.~262--273, 2013.

\bibitem{ma2016virtual}
J.~Ma and F.~Zhou, ``Virtual dynamic message signs: a future mode for basic
  public traveller information,'' {\em IET Intelligent Transport Systems},
  vol.~10, no.~7, pp.~476--482, 2016.

\bibitem{chmiel2016insigma}
W.~Chmiel, J.~Da{\'n}da, A.~Dziech, S.~Ernst, P.~Kad{\l}uczka, Z.~Mikrut,
  P.~Pawlik, P.~Szwed, and I.~Wojnicki, ``Insigma: an intelligent
  transportation system for urban mobility enhancement,'' {\em Multimedia Tools
  and Applications}, vol.~75, no.~17, pp.~10529--10560, 2016.

\bibitem{darwish2017empowering}
S.~M. Darwish, ``Empowering vehicle tracking in a cluttered environment with
  adaptive cellular automata suitable to intelligent transportation systems,''
  {\em IET Intelligent Transport Systems}, vol.~11, no.~2, pp.~84--91, 2017.

\bibitem{altinkaya2013urban}
M.~Altinkaya and M.~Zontul, ``Urban bus arrival time prediction: A review of
  computational models,'' {\em International Journal of Recent Technology and
  Engineering (IJRTE)}, vol.~2, no.~4, pp.~164--169, 2013.

\bibitem{mukheja2017smartphone}
P.~Mukheja, N.~R. Velaga, R.~Sharmila, {\em et~al.}, ``Smartphone-based
  crowdsourcing for position estimation of public transport vehicles,'' {\em
  IET Intelligent Transport Systems}, vol.~11, no.~9, pp.~588--595, 2017.

\bibitem{zuazola2013telematics}
I.~J.~G. Zuazola, A.~Moreno, H.~Landaluce, I.~Angulo, A.~Perallos,
  U.~Hern{\'a}ndez-Jayo, N.~Sainz, M.~A. Ziai, J.~C. Batchelor, and
  J.~Elmirghani, ``Telematics system for the intelligent transport and
  distribution of medicines,'' {\em IET Intelligent Transport Systems}, vol.~7,
  no.~1, pp.~131--137, 2013.

\bibitem{hossinger2014development}
R.~H{\"o}ssinger, P.~Widhalm, M.~Ulm, K.~Heimbuchner, E.~Wolf, R.~Apel, and
  T.~Uhlmann, ``Development of a real-time model of the occupancy of short-term
  parking zones,'' {\em International Journal of Intelligent Transportation
  Systems Research}, vol.~12, no.~2, pp.~37--47, 2014.

\bibitem{coulibaly2018development}
M.~Coulibaly, S.~Belkhala, A.~Errami, H.~Medromi, A.~Saad, M.~Rouissiya, and
  A.~Jaafari, ``Development of a demonstrator
  {\guillemotleft}smart-parking{\guillemotright},'' in {\em Proc. of the 19th
  IEEE Mediterranean Electrotechnical Conference (MELECON)}, pp.~172--176,
  IEEE, 2018.

\bibitem{smartilabsmartypark}
Smartilab, ``Smartypark.'' \url{http://www.smartilab.ma/smartypark/}.

\bibitem{Policyvictoria}
T.~Litman, ``Autonomous vehicle implementation predictions implications for
  transport planning,'' tech. rep., Victoria Transport Policy Institute, 2019.

\bibitem{Bagloee2016}
S.~A. Bagloee, M.~Tavana, M.~Asadi, and T.~Oliver, ``Autonomous vehicles:
  challenges, opportunities, and future implications for transportation
  policies,'' {\em Journal of Modern Transportation}, vol.~24, pp.~284--303,
  Dec 2016.

\bibitem{zeng2019joint}
T.~Zeng, O.~Semiari, W.~Saad, and M.~Bennis, ``Joint communication and control
  system design for connected and autonomous vehicle navigation,'' in {\em
  Proc. of the IEEE International Conference on Communications ICC 2019-2019},
  pp.~1--6, IEEE, 2019.

\bibitem{brainonboard}
``Safety features: Driver assistance technology.''
  \url{{http://brainonboard.ca/safety_features/driver_assistance_technology.php}}.

\bibitem{liu2018urban}
Z.~Liu, Z.~Li, K.~Wu, and M.~Li, ``Urban traffic prediction from mobility data
  using deep learning,'' {\em IEEE Network}, vol.~32, no.~4, pp.~40--46, 2018.

\bibitem{chen2019artificial}
M.~Chen, U.~Challita, W.~Saad, C.~Yin, and M.~Debbah, ``Artificial neural
  networks-based machine learning for wireless networks: A tutorial,'' {\em
  IEEE Communications Surveys \& Tutorials}, vol.~21, no.~4, pp.~3039--3071,
  2019.

\bibitem{jan2017deep}
B.~Jan, H.~Farman, M.~Khan, M.~Imran, I.~U. Islam, A.~Ahmad, S.~Ali, and
  G.~Jeon, ``Deep learning in big data analytics: A comparative study,'' {\em
  Computers \& Electrical Engineering}, 2017.

\bibitem{lv2018lc}
Z.~Lv, J.~Xu, K.~Zheng, H.~Yin, P.~Zhao, and X.~Zhou, ``Lc-rnn: A deep learning
  model for traffic speed prediction.,'' in {\em IJCAI}, pp.~3470--3476, 2018.

\bibitem{yao2019revisiting}
H.~Yao, X.~Tang, H.~Wei, G.~Zheng, and Z.~Li, ``Revisiting spatial-temporal
  similarity: A deep learning framework for traffic prediction,'' in {\em Proc.
  of the AAAI Conference on Artificial Intelligence}, 2019.

\bibitem{essien2019improving}
A.~Essien, I.~Petrounias, P.~Sampaio, and S.~Sampaio, ``Improving urban traffic
  speed prediction using data source fusion and deep learning,'' in {\em Proc.
  of the IEEE International Conference on Big Data and Smart Computing
  (BigComp)}, pp.~1--8, IEEE, 2019.

\bibitem{zonoozi2018periodic}
A.~Zonoozi, J.-j. Kim, X.-L. Li, and G.~Cong, ``Periodic-crn: A convolutional
  recurrent model for crowd density prediction with recurring periodic
  patterns.,'' in {\em IJCAI}, pp.~3732--3738, 2018.

\bibitem{cheng2018deeptransport}
X.~Cheng, R.~Zhang, J.~Zhou, and W.~Xu, ``Deeptransport: Learning
  spatial-temporal dependency for traffic condition forecasting,'' in {\em
  Proc. of the International Joint Conference on Neural Networks (IJCNN)},
  pp.~1--8, IEEE, 2018.

\bibitem{wang2018deepstcl}
D.~Wang, Y.~Yang, and S.~Ning, ``Deepstcl: A deep spatio-temporal convlstm for
  travel demand prediction,'' in {\em Proc. of the International Joint
  Conference on Neural Networks (IJCNN)}, pp.~1--8, IEEE, 2018.

\bibitem{chen2016learning}
Q.~Chen, X.~Song, H.~Yamada, and R.~Shibasaki, ``Learning deep representation
  from big and heterogeneous data for traffic accident inference,'' in {\em
  Proc. of the thirtieth AAAI Conference on Artificial Intelligence}, 2016.

\bibitem{lamr2018big}
M.~Lamr, ``Big data and its usage in systems of early warning of traffic
  accident risks,'' in {\em Proc. of the Sixth International Conference on
  Enterprise Systems (ES)}, pp.~154--157, IEEE, 2018.

\bibitem{bhattacharya2018eye}
S.~Bhattacharya and S.~Bernadin, ``Eye-glance frequency as a function of
  driver's intent to change lanes,'' in {\em Proc. of the 88th IEEE Vehicular
  Technology Conference (VTC-Fall)}, pp.~1--5, IEEE, 2018.

\bibitem{cao2018multiagent}
Z.~Cao, H.~Guo, and J.~Zhang, ``A multiagent-based approach for vehicle routing
  by considering both arriving on time and total travel time,'' {\em ACM
  Transactions on Intelligent Systems and Technology (TIST)}, vol.~9, no.~3,
  p.~25, 2018.

\bibitem{cao2017maximizing}
Z.~Cao, H.~Guo, J.~Zhang, F.~Oliehoek, and U.~Fastenrath, ``Maximizing the
  probability of arriving on time: A practical q-learning method,'' in {\em
  Proc. of the thirty-First AAAI Conference on Artificial Intelligence}, 2017.

\bibitem{cao2016multiagent}
Z.~Cao, H.~Guo, J.~Zhang, and U.~Fastenrath, ``Multiagent-based route guidance
  for increasing the chance of arrival on time,'' in {\em Proc. of the
  thirtieth AAAI Conference on Artificial Intelligence}, 2016.

\bibitem{cao2015finding}
Z.~Cao, H.~Guo, J.~Zhang, D.~Niyato, and U.~Fastenrath, ``Finding the shortest
  path in stochastic vehicle routing: A cardinality minimization approach,''
  {\em IEEE Transactions on Intelligent Transportation Systems}, vol.~17,
  no.~6, pp.~1688--1702, 2015.

\bibitem{cao2016unified}
Z.~Cao, S.~Jiang, J.~Zhang, and H.~Guo, ``A unified framework for vehicle
  rerouting and traffic light control to reduce traffic congestion,'' {\em IEEE
  transactions on intelligent transportation systems}, vol.~18, no.~7,
  pp.~1958--1973, 2016.

\bibitem{guo2017routing}
H.~Guo, Z.~Cao, M.~Seshadri, J.~Zhang, D.~Niyato, and U.~Fastenrath, ``Routing
  multiple vehicles cooperatively: Minimizing road network breakdown
  probability,'' {\em IEEE Transactions on Emerging Topics in Computational
  Intelligence}, vol.~1, no.~2, pp.~112--124, 2017.

\bibitem{martinez2017energy}
C.~M. Martinez, X.~Hu, D.~Cao, E.~Velenis, B.~Gao, and M.~Wellers, ``Energy
  management in plug-in hybrid electric vehicles: Recent progress and a
  connected vehicles perspective,'' 2017.

\bibitem{Efthymiou2017}
D.~Efthymiou, K.~Chrysostomou, M.~Morfoulaki, and G.~Aifantopoulou, ``Electric
  vehicles charging infrastructure location: a genetic algorithm approach,''
  {\em European Transport Research Review}, vol.~9, p.~27, May 2017.

\bibitem{elbanhawy2014investigating}
E.~Y. ElBanhawy, ``Investigating the evolution of e-mobility in its urban
  context a longitudinal study in newcastle-gateshead area,'' 2014.

\bibitem{fatnassi2015viability}
E.~Fatnassi, O.~Chebbi, and J.~Chaouachi, ``Viability of implementing smart
  mobility tool in the case of tunis city,'' in {\em Proc. of the IFIP
  International Conference on Computer Information Systems and Industrial
  Management}, pp.~339--350, Springer, 2015.

\bibitem{arena2014service}
M.~Arena, G.~Azzone, A.~Colorni, A.~Conte, A.~Lu{\`e}, and R.~Nocerino,
  ``Service design in electric vehicle sharing: evidence from italy,'' {\em IET
  Intelligent Transport Systems}, vol.~9, no.~2, pp.~145--155, 2014.

\bibitem{marrakechelectricbus}
C.~de~Marrakech, ``Lancement des bus electriques.''
  \url{http://www.ville-marrakech.ma/lancement-des-bus-electriques/996/}.

\bibitem{guidecasatransportcasa}
C.~Transport, ``Le guide pour voyager smart dans une ville qui respire,'' tech.
  rep.

\bibitem{yang2017ev}
T.~Yang, X.~Xu, Q.~Guo, L.~Zhang, and H.~Sun, ``Ev charging behaviour analysis
  and modelling based on mobile crowdsensing data,'' {\em IET Generation,
  Transmission \& Distribution}, vol.~11, no.~7, pp.~1683--1691, 2017.

\bibitem{sohet2020coupled}
B.~Sohet, Y.~Hayel, O.~Beaude, and A.~Jeandin, ``Coupled charging-and-driving
  incentives design for electric vehicles in urban networks,'' {\em arXiv
  preprint arXiv:2001.11758}, 2020.

\bibitem{ma2011decentralized}
Z.~Ma, D.~S. Callaway, and I.~A. Hiskens, ``Decentralized charging control of
  large populations of plug-in electric vehicles,'' {\em IEEE Transactions on
  control systems technology}, vol.~21, no.~1, pp.~67--78, 2011.

\bibitem{etesami2017smart}
S.~R. Etesami, W.~Saad, N.~Mandayam, and H.~V. Poor, ``Smart routing in smart
  grids,'' in {\em Proc. of the 56th IEEE Annual Conference on Decision and
  Control (CDC)}, pp.~2599--2604, IEEE, 2017.

\bibitem{obama2009memorandum}
B.~Obama, ``Memorandum for the heads of executive departments and agencies,''
  {\em Presidential Studies Quarterly}, vol.~39, no.~3, p.~429, 2009.

\bibitem{manyika2013open}
J.~Manyika, M.~Chui, P.~Groves, D.~Farrell, S.~Van~Kuiken, and E.~A. Doshi,
  ``Open data: Unlocking innovation and performance with liquid information,''
  {\em McKinsey Global Institute}, vol.~21, 2013.

\bibitem{carrara2015creating}
W.~Carrara, W.~Chan, S.~Fischer, and E.~v. Steenbergen, ``Creating value
  through open data: Study on the impact of re-use of public data resources,''
  {\em European Commission}, 2015.

\bibitem{TransportforLondon}
T.~for London, ``Open data users.''
  \url{https://tfl.gov.uk/info-for/open-data-users/}.

\bibitem{opendatacensus}
O.~K. Foundation, ``Local for open data census.''
  \url{https://census.okfn.org/en/latest/local/}.

\bibitem{tyrinopoulos2013factors}
Y.~Tyrinopoulos and C.~Antoniou, ``Factors affecting modal choice in urban
  mobility,'' {\em European Transport Research Review}, vol.~5, no.~1, p.~27,
  2013.

\bibitem{igidatareliability}
I.~G.~D. of~Knowledge, ``What is data reliability.''
  \url{https://www.igi-global.com/dictionary/data-reliability/6801}, Jan. 29,
  2019.

\bibitem{ebrarydatacollection}
Ebrary, ``Considerations for collecting data.''
  \url{https://ebrary.net/1291/education/considerations_for_collecting_data}.

\bibitem{igidatafreshness}
I.~G.~D. of~Knowledge, ``What is data freshness.''
  \url{https://www.igi-global.com/dictionary/data-freshness/6731}, Jan. 29,
  2019.

\bibitem{igidataattribute}
I.~G.~D. of~Knowledge, ``What is data quality attribute.''
  \url{https://www.igi-global.com/dictionary/data-quality-attribute/6793}, Jan.
  29, 2019.

\bibitem{zhou2019joint}
B.~Zhou and W.~Saad, ``Joint status sampling and updating for minimizing age of
  information in the internet of things,'' {\em IEEE Transactions on
  Communications}, vol.~67, no.~11, pp.~7468--7482, 2019.

\bibitem{wang2016travel}
Y.~Wang, B.~Du, Q.~Rong, and X.~Lin, ``Travel patterns analysis of urban
  residents using automated fare collection system,'' {\em Chinese Journal of
  Electronics}, vol.~25, no.~1, pp.~40--47, 2016.

\bibitem{jung2017deep}
J.~Jung and K.~Sohn, ``Deep-learning architecture to forecast destinations of
  bus passengers from entry-only smart-card data,'' {\em IET Intelligent
  Transport Systems}, vol.~11, no.~6, pp.~334--339, 2017.

\bibitem{wan2016predictability}
S.~Wan, J.~Meng, S.~Fang, X.~Xing, K.~Xie, and K.~Bian, ``Predictability
  analysis on expressway vehicle mobility using electronic toll collection
  data,'' in {\em Proc. of the 19th IEEE International Conference on
  Intelligent Transportation Systems (ITSC)}, pp.~2589--2594, IEEE, 2016.

\bibitem{fan2018using}
S.-K.~S. Fan, C.-J. Su, H.-T. Nien, P.-F. Tsai, and C.-Y. Cheng, ``Using
  machine learning and big data approaches to predict travel time based on
  historical and real-time data from taiwan electronic toll collection,'' {\em
  Soft Computing}, vol.~22, no.~17, pp.~5707--5718, 2018.

\bibitem{schneider2009traveller}
M.~Schneider, M.~Linauer, N.~Hainitz, and H.~Koller, ``Traveller information
  service based on real-time toll data in austria,'' {\em IET Intelligent
  Transport Systems}, vol.~3, no.~2, pp.~124--137, 2009.

\bibitem{christopher2019smart}
K.~K. Christopher, X.~V. Arul, and P.~Karthikeyen, ``Smart toll tax automation
  and monitoring system using android application,'' in {\em Proc. of the IEEE
  International Conference on Intelligent Techniques in Control, Optimization
  and Signal Processing (INCOS)}, pp.~1--6, IEEE, 2019.

\bibitem{ghosh2017intelligent}
R.~Ghosh, R.~Pragathi, S.~Ullas, and S.~Borra, ``Intelligent transportation
  systems: A survey,'' in {\em Proc. of 2017 International Conference on
  Circuits, Controls, and Communications (CCUBE)}, pp.~160--165, IEEE, 2017.

\bibitem{castro2013taxi}
P.~S. Castro, D.~Zhang, C.~Chen, S.~Li, and G.~Pan, ``From taxi gps traces to
  social and community dynamics: A survey,'' {\em ACM Computing Surveys
  (CSUR)}, vol.~46, no.~2, p.~17, 2013.

\bibitem{calabrese2015urban}
F.~Calabrese, L.~Ferrari, and V.~D. Blondel, ``Urban sensing using mobile phone
  network data: a survey of research,'' {\em ACM Computing Surveys}, vol.~47,
  no.~2, p.~25, 2015.

\bibitem{Tosi2017}
D.~Tosi, ``Cell phone big data to compute mobility scenarios for future smart
  cities,'' {\em International Journal of Data Science and Analytics}, vol.~4,
  pp.~265--284, Dec 2017.

\bibitem{liu2016data}
J.~Liu, L.~Bic, H.~Gong, and S.~Zhan, ``Data collection for mobile crowdsensing
  in the presence of selfishness,'' {\em EURASIP journal on wireless
  communications and networking}, vol.~2016, no.~1, p.~82, 2016.

\bibitem{stojanovic2016mobile}
D.~Stojanovic, B.~Predic, and N.~Stojanovic, ``Mobile crowd sensing for smart
  urban mobility,'' {\em European Handbook of Crowdsourced Geographic
  Information}, pp.~371--382, 2016.

\bibitem{meseguer2017drivingstyles}
J.~E. Meseguer, C.~K. Toh, C.~T. Calafate, J.~C. Cano, and P.~Manzoni,
  ``Drivingstyles: a mobile platform for driving styles and fuel consumption
  characterization,'' {\em Journal of Communications and networks}, vol.~19,
  no.~2, pp.~162--168, 2017.

\bibitem{assem2017rcmc}
H.~Assem, T.~S. Buda, and D.~O'sullivan, ``Rcmc: recognizing crowd-mobility
  patterns in cities based on location based social networks data,'' {\em ACM
  Transactions on Intelligent Systems and Technology (TIST)}, vol.~8, no.~5,
  p.~70, 2017.

\bibitem{sharma2017sustainable}
P.~Sharma and S.~Rajput, {\em Sustainable smart cities in India: Challenges and
  future perspectives}.
\newblock Springer, 2017.

\bibitem{zhang2018real}
D.~Zhang, T.~He, and F.~Zhang, ``Real-time human mobility modeling with
  multi-view learning,'' {\em ACM Transactions on Intelligent Systems and
  Technology (TIST)}, vol.~9, no.~3, p.~22, 2018.

\bibitem{long2019predicting}
K.~Long, W.~Yao, J.~Gu, W.~Wu, and L.~D. Han, ``Predicting freeway travel time
  using multiple-source heterogeneous data integration,'' {\em Applied
  Sciences}, vol.~9, no.~1, p.~104, 2019.

\bibitem{zheng2014urban}
Y.~Zheng, L.~Capra, O.~Wolfson, and H.~Yang, ``Urban computing: concepts,
  methodologies, and applications,'' {\em ACM Transactions on Intelligent
  Systems and Technology (TIST)}, vol.~5, no.~3, p.~38, 2014.

\bibitem{wang2017computing}
S.~Wang, X.~Zhang, J.~Cao, L.~He, L.~Stenneth, P.~S. Yu, Z.~Li, and Z.~Huang,
  ``Computing urban traffic congestions by incorporating sparse gps probe data
  and social media data,'' {\em ACM Transactions on Information Systems
  (TOIS)}, vol.~35, no.~4, p.~40, 2017.

\bibitem{manolopoulos2013smartphone}
V.~Manolopoulos, S.~Tao, A.~Rusu, and P.~Papadimitratos, ``Smartphone-based
  traffic information system for sustainable cities,'' {\em ACM SIGMOBILE
  Mobile Computing and Communications Review}, vol.~16, no.~4, pp.~30--31,
  2013.

\bibitem{morocco2014programforresultstransport}
Morocco, ``Program-for-results information document (pid) concept stage,''
  2014.

\bibitem{morocco2015urbantransportprogram}
Morocco, ``Program-for-results information document (pid) appraisal stage,''
  2015.

\bibitem{top50}
E.~S. Institute, ``Top 50 smart city governments,'' tech. rep.

\bibitem{nesh2015citizen}
T.~Nesh-Nash and Z.~Mahrez, ``Social partnership for effective citizen
  engagement,'' {\em Ecocities in challenging environments}, 2015.

\bibitem{ross2015hyperloop}
P.~E. Ross, ``Hyperloop: no pressure,'' {\em IEEE Spectrum}, vol.~53, no.~1,
  pp.~51--54, 2015.

\bibitem{behruz2013challenges}
H.~Behruz, A.~Chavoshy, G.~Mozaffari, {\em et~al.}, ``Challenges of
  implementation of intelligent transportation systems in developing countries:
  case study--tehran,'' {\em WIT Transactions on Ecology and the Environment},
  vol.~179, pp.~977--987, 2013.

\bibitem{mozaffari2019tutorial}
M.~Mozaffari, W.~Saad, M.~Bennis, Y.-H. Nam, and M.~Debbah, ``A tutorial on
  uavs for wireless networks: Applications, challenges, and open problems,''
  {\em IEEE communications surveys \& tutorials}, vol.~21, no.~3,
  pp.~2334--2360, 2019.

\bibitem{chen2017caching}
M.~Chen, M.~Mozaffari, W.~Saad, C.~Yin, M.~Debbah, and C.~S. Hong, ``Caching in
  the sky: Proactive deployment of cache-enabled unmanned aerial vehicles for
  optimized quality-of-experience,'' {\em IEEE Journal on Selected Areas in
  Communications}, vol.~35, no.~5, pp.~1046--1061, 2017.

\bibitem{mozaffari2017mobile}
M.~Mozaffari, W.~Saad, M.~Bennis, and M.~Debbah, ``Mobile unmanned aerial
  vehicles (uavs) for energy-efficient internet of things communications,''
  {\em IEEE Transactions on Wireless Communications}, vol.~16, no.~11,
  pp.~7574--7589, 2017.

\bibitem{ferdowsi2019cyber}
A.~Ferdowsi, S.~Ali, W.~Saad, and N.~B. Mandayam, ``Cyber-physical security and
  safety of autonomous connected vehicles: optimal control meets multi-armed
  bandit learning,'' {\em IEEE Transactions on Communications}, vol.~67,
  no.~10, pp.~7228--7244, 2019.

\bibitem{huqcyberattacks}
N.~Huq, R.~Vosseler, and M.~Swimmer, ``Cyberattacks against intelligent
  transportation systems,''

\bibitem{martylegal}
A.~MARTY, ``Legal and ethical considerations in the era of autonomous robots,''

\end{thebibliography}

\end{document}